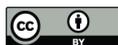

# Dawn–dusk asymmetries in the coupled solar wind–magnetosphere–ionosphere system: a review

A. P. Walsh[1], S. Haaland[2,3], C. Forsyth[4], A. M. Keesee[5], J. Kissinger[6], K. Li[2], A. Runov[7], J. Soucek[8], B. M. Walsh[6,11], S. Wing[9], and M. G. G. T. Taylor[10]

[1]Science and Robotic Exploration Directorate, European Space Agency, ESAC, Villanueva de la Cañada, Madrid, Spain
[2]Max-Planck-Institute for Solar System Research, Göttingen, Germany
[3]Birkeland Center for Space Science, University of Bergen, Bergen, Norway
[4]UCL Department of Space and Climate Physics, Mullard Space Science Laboratory, Holmbury St. Mary, Surrey, UK
[5]West Virginia University, Morgantown, West Virginia, USA
[6]NASA Goddard Space Flight Center, Greenbelt, Maryland, USA
[7]Department of Earth and Space Sciences, UCLA, Los Angeles, California, USA
[8]Institute of Atmospheric Physics, Czech Academy of Sciences, Prague, Czech Republic
[9]Johns Hopkins University Applied Physics Laboratory, Maryland, USA
[10]Science and Robotic Exploration Directorate, European Space Agency, ESTEC, Noordwijk, The Netherlands
[11]Space Sciences Laboratory, University of California, Berkeley, USA

*Correspondence to:* A. P. Walsh (andrew.walsh@esa.int)

**Abstract.** Dawn–dusk asymmetries are ubiquitous features of the coupled solar-wind–magnetosphere–ionosphere system. During the last decades, increasing availability of satellite and ground-based measurements has made it possible to study these phenomena in more detail. Numerous publications have documented the existence of persistent asymmetries in processes, properties and topology of plasma structures in various regions of geospace. In this paper, we present a review of our present knowledge of some of the most pronounced dawn–dusk asymmetries. We focus on four key aspects: (1) the role of external influences such as the solar wind and its interaction with the Earth's magnetosphere; (2) properties of the magnetosphere itself; (3) the role of the ionosphere and (4) feedback and coupling between regions. We have also identified potential inconsistencies and gaps in our understanding of dawn–dusk asymmetries in the Earth's magnetosphere and ionosphere.

**Keywords.** Magnetospheric physics (magnetosphere–ionosphere interactions; magnetospheric configuration and dynamics; solar-wind–magnetosphere interactions)

## 1 Introduction

In recent years, increasing availability of remotely sensed and in situ measurements of the ionosphere, magnetosphere and magnetosheath have allowed ever-larger statistical studies to be carried out. Equally, advances in technology and methodology have allowed increasingly detailed and realistic simulations. These studies and simulations have revealed significant, persistent dawn–dusk asymmetries throughout the solar-wind–magnetosphere–ionosphere system. Dawn–dusk asymmetries have been observed in the Earth's magnetotail current systems and particle fluxes; in the ring current; and in polar cap patches and the global convection pattern in the ionosphere. Various authors have related these asymmetries to differences in solar illumination, ionospheric conductivity and processes internal to the magnetosphere. Significant dawn–dusk asymmetries have also been observed in the terrestrial magnetosheath, and there is evidence that plasma entry mechanisms to the magnetotail, for example, operate differently in the pre- and post-midnight sectors.

The purpose of this review is to identify and collect current knowledge about dawn–dusk asymmetries, examining the solar-wind–magnetosphere–ionosphere system as a whole.

We consider the roles that coupling between the solar wind and magnetosphere, between the magnetosphere and ionosphere, and between different plasma regimes within the magnetosphere itself play in creating and supporting these asymmetries. We provide a schematic summary of current understanding of dawn–dusk asymmetries (Fig. 18), and also highlight inconsistencies and gaps in this knowledge, identifying possible directions for future work in this area.

## 2 Observed asymmetries

In this section we review the various dawn–dusk asymmetries that have been observed in the solar-wind–magnetosphere–ionosphere system.

### 2.1 Solar wind and interplanetary magnetic field

The outer layers of geospace, from the foreshock inward through the magnetosheath to the magnetopause, are formed from the incident solar wind perturbed by the terrestrial magnetic field. A number of dawn–dusk asymmetries arise in these regions. The first asymmetry comes from the orbital motion of the Earth around the Sun. This motion causes the direction of the solar wind flow in a geocentric reference frame to be aberrated from the Earth–Sun line by roughly four degrees for a typical solar wind velocity. This provides a natural axis of symmetry for studies of dawn–dusk asymmetries in the magnetospheric system and is often called an "aberrated" coordinate system.

The second upstream asymmetry comes from the average orientation of the interplanetary magnetic field (IMF) permeating the solar wind. The IMF vector is variable, but the average orientation follows the Parker spiral. Since the direction is typically not aligned with the solar wind flow, an asymmetry is introduced to the magnetospheric system due to a different orientation of IMF with respect to the bow shock normal in the dawn and dusk sectors. Figure 1 shows the average properties of the IMF; the two maxima in the $B_X$ and $B_Y$ histogram correspond to the inward and outward Parker spiral orientation.

#### 2.1.1 Foreshock

The foreshock is the region of the solar wind magnetically connected to the bow shock. Its geometry, properties and location are mediated by the IMF. Under the typical Parker spiral IMF, the foreshock is formed on the dawn side, where the angle between IMF and the shock normal ($\Theta_{Bn}$) is small and the particles can more easily cross the shock front. Since the IMF and bow shock normal vector are close to parallel this region is called the quasi-parallel shock, as opposed to the quasi-perpendicular shock, where IMF is nearly tangent to the shock surface and the foreshock is not formed. The generation of the foreshock therefore provides an upstream "boundary condition" for magnetosheath processes that vary between the dawn and dusk sides.

The foreshock differs from the pristine unperturbed solar wind by the presence of particles (electrons and ions) back-streaming away from the shock. These particles are responsible for the generation of various kinds of the foreshock plasma. Both the particles and plasma oscillations can be convected back to the shock and drive shock or magnetosheath oscillation. A detailed review of foreshock properties can be found in Eastwood et al. (2005b); here we review only aspects relevant to asymmetries induced farther downstream.

The foreshock region is conventionally divided into two parts – electron and ion. The electron foreshock, the upstream-most part adjacent to the IMF line tangent to the shock, populated by back-streaming electrons only and associated electron plasma waves (Filbert and Kellogg, 1979). The processes in the electron foreshock have very little influence on the shock and the magnetosheath. On the other hand, the processes in the ion foreshock, where reflected and back-streaming ions are also present (Meziane et al., 2004), influence the bow shock and the magnetosheath significantly.

Figure 2 shows the geometry and magnetic field configuration of the ion foreshock, bow shock and magnetosheath. The distribution function plots show the diffuse hot ions leaking from the quasi-parallel shock back into the solar wind (Gosling et al., 1989). The ultra-low frequency (ULF) waves in the ion foreshock were identified as fast-mode magnetosonic waves, generated by the ion beams (Archer et al., 2005; Eastwood et al., 2005a). Note that the region populated by waves is a sub-section of the ion foreshock, separated by a clear boundary, called the foreshock compressional boundary (e.g. Omidi et al., 2009).

The foreshock ULF waves are typically propagating upstream in the plasma rest frame, but are convected downstream by the solar wind and enter the quasi-parallel shock region, modulating the shock (Sibeck and Gosling, 1996) and possibly being transmitted in the magnetosheath (Engebretson et al., 1991), as discussed in Sect. 2.1.2. Since the foreshock only occupies the area upstream of the quasi-parallel shock, this transmission of foreshock oscillations in the magnetosheath only occurs on the quasi-parallel side of the magnetosheath (dawn side for Parker spiral IMF orientation), introducing a dawn–dusk asymmetry into the magnetosheath.

#### 2.1.2 Magnetosheath asymmetries

Standing fast-mode waves known as bow shocks decelerate and deflect the supersonic and super Alfvénic solar wind, enabling it to pass around planetary and cometary obstacles throughout the heliosphere. The transition region between a bow shock and its obstacle is called the magnetosheath. Early theoretical considerations proposed dawn–dusk asymmetries of density, temperature, pressure and bulk flow within the magnetosheath (Walters, 1964). These predictions were based on differing Rankine–Hugoniot shock jump conditions with a magnetic field parallel or perpendicular to the bow shock. A Parker spiral magnetic configuration incident upon the bow shock would introduce the necessary geometry for dawn–dusk asymmetries.

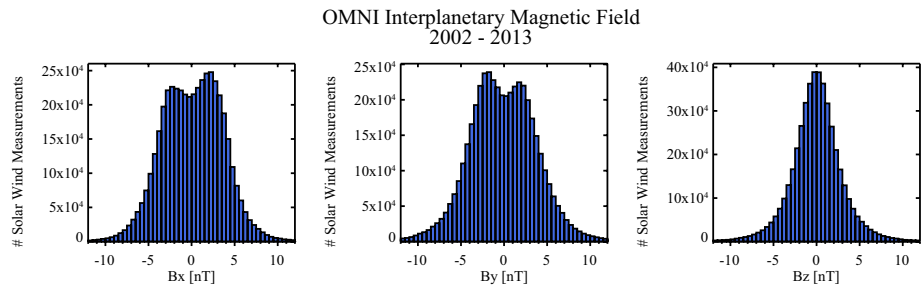

**Figure 1.** Histograms of interplanetary magnetic field built from 1 min OMNI data over one solar cycle (January 2002 to August 2013). Each panel shows a histogram of one IMF component in the GSE coordinate system. The two maxima in the $B_X$ and $B_Y$ plots correspond to inward and outward Parker spiral direction, the most probable IMF orientation.

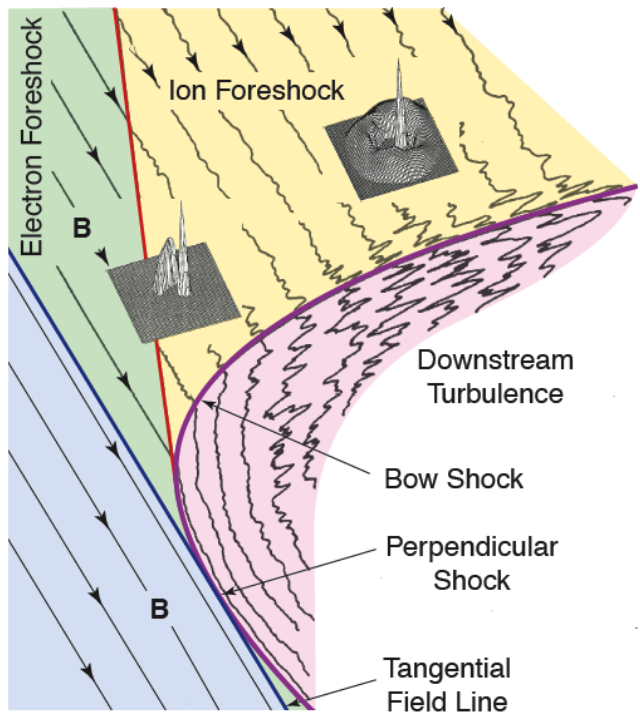

**Figure 2.** Schematic view of the foreshock, bow shock and magnetosheath of the Earth. The ripples in the magnetic field represent foreshock ULF waves and turbulence downstream of quasi-parallel shock. Distribution function plots show the field-aligned ion beams (close to the ion foreshock boundary) and the diffuse (close to the quasi-parallel shock) ions. Adapted from Balogh and Treumann (2013).

Since these early theoretical predictions, a number of statistical studies have been conducted with a variety of spacecraft and have found a range of asymmetries in the magnetosheath (see summary in Table 1). One parameter that has been studied by a number of authors is the ion plasma density. Although higher ion density was observed in the dawn magnetosheath through a number of studies, the magnitude of this asymmetry varied from 1 to 33 %. Several studies proposed an IMF source of the asymmetry, but were unable to confirm this through binning the measurements by upstream IMF (Paularena et al., 2001; Longmore et al., 2005). One possible reason for this result is the limited statistics available for ortho-Parker spiral IMF, or an IMF when the quasi-parallel bow shock is on the duskside.

Walsh et al. (2012) proposed that the density asymmetry resulted from an asymmetric bow shock shape in response to the direction of the IMF. The bow shock is a fast-mode wave, which travels faster perpendicular to a magnetic field than parallel to it (Wu, 1992; Chapman et al., 2004). This results in a bow shock that is radially farther from the Earth on the duskside than the dawn when the IMF is in a Parker spiral orientation. Figure 3 shows the impact of the IMF angle on the bow shock position and Alfvénic Mach number through magnetohydrodynamics (MHD). An additional feature shown in the figure is that the asymmetry is a function of the Alfvénic Mach number. Since the average Alfvénic Mach number in the solar wind varies with the phase of the solar cycle (Luhmann et al., 1993), the magnitude of the density asymmetry in the average magnetosheath should also vary with phase of the solar cycle (larger asymmetry during solar minimum). Walsh et al. (2012) looked at the average Alfvénic Mach number during each of the past studies and found good agreement with the expected trend in the density asymmetry. An asymmetric bow shock position resulting from the Parker spiral IMF also explains the asymmetries observed in ion temperature and magnetic field (see Table 1).

### 2.1.3 Waves and kinetic effects in the magnetosheath

In addition to asymmetries in plasma moments and magnetic field magnitude in the magnetosheath, there are also observed asymmetries in the waves and kinetic effects. Since the first spacecraft observations, it has been known that the magnetosheath is populated by turbulent field and plasma oscillations covering the frequency range from the timescale of minutes to well above the ion plasma frequency. Early works suggested that magnetic field fluctuations can originate both from the upstream solar wind and foreshock, as well as from the magnetopause, while some are generated by plasma instabilities within the magnetosheath itself (for a review, see Fairfield, 1976).

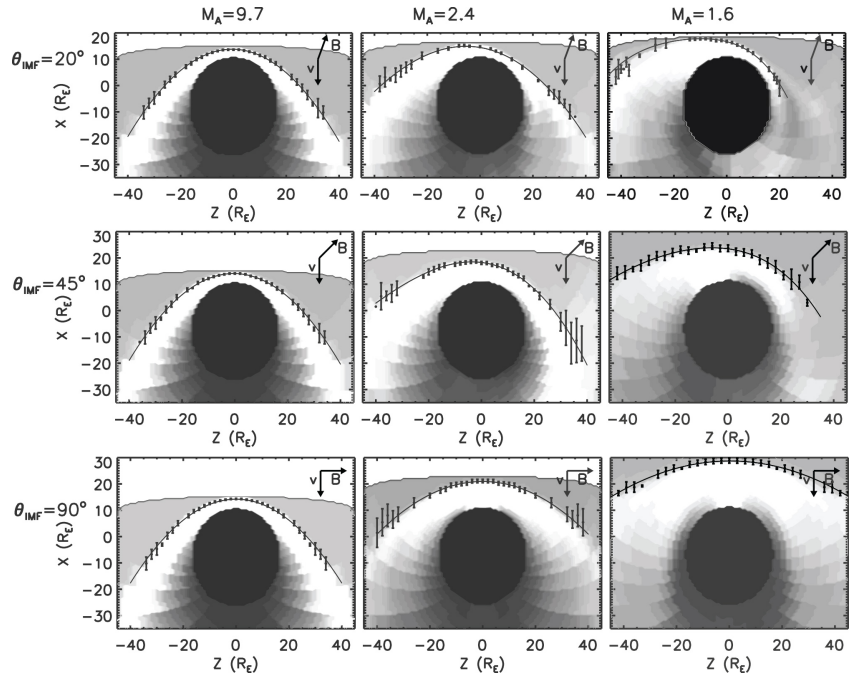

**Figure 3.** Adapted from Chapman et al. (2004). The bow shock position and plasma density is shown from MHD simulations with varying Alfvénic Mach number and magnetic field orientation. From left to right the Alfvénic Mach number decreases. From top to bottom the orientation of the magnetic field changes from close to parallel to the flow direction to 90° from it.

Fairfield and Ness (1970) noted a dawn–dusk asymmetry in the amplitude of magnetic field oscillations. Later systematic studies with the aid of an upstream solar wind monitor have established that the IMF $B_Y$ component and consequently the $\Theta_{Bn}$ parameter of the upstream shock are important factors in determining the properties of magnetosheath fluctuations. Luhmann et al. (1986) demonstrated an increased level of magnetosheath field fluctuations (using 4 s resolution data) behind the quasi-parallel shock. Two decades later, Shevyrev et al. (2007) showed that the direction of the field varied much more in the quasi-parallel magnetosheath than in the quasi-perpendicular. This effect is visualised in Fig. 4 adapted from Petrinec (2013), who presented a global view of magnetosheath field fluctuations using median magnetic field measurements from Geotail observations, restricted to Parker spiral IMF direction.

The above studies confirmed that the quasi-parallel shock is a more efficient source of magnetosheath oscillations at longer timescales (wave periods > 1 min) and that the oscillations resemble solar wind turbulence. Controversy remains concerning the precise generating mechanism of the turbulence at the quasi-parallel shock. Locally generated turbulence at the shock (Greenstadt et al., 2001; Luhmann et al., 1986) and transmission of upstream foreshock fluctuations (Engebretson et al., 1991; Sibeck and Gosling, 1996; Němeček et al., 2002) were proposed. Gutynska et al. (2012) investigated multi-spacecraft correlations between the magnetosheath and solar wind and concluded that fluctuations with wave periods larger than 100 s can often be traced back to solar wind fluctuations, while smaller-scale fluctuations are not correlated with upstream waves.

Consistent with this result, field and plasma oscillations in the quasi-perpendicular magnetosheath are typically smaller in amplitude and more compressive in nature (e.g. Shevyrev et al., 2007). This can be explained by the dominance of locally generated kinetic waves and, most importantly, mirror modes. Magnetosheath ions are characterised by relatively high $\beta$ (> 1) and significant temperature anisotropy $T_\perp/T_\parallel > 1$, giving rise to two kinetic instabilities – ion cyclotron instability and mirror instability. In the magnetosheath plasma, these two instabilities often compete and both modes are frequently observed (for a review, see Schwartz et al., 1996; Lucek et al., 2005). These waves typically appear at shorter timescales, below one minute, and can grow to significant amplitudes.

Anderson and Fuselier (1993) compared the occurrence rates of mirror and EMIC waves for quasi-perpendicular and quasi-parallel shock conditions. Wave character was identified by spectral analysis and the nature of the shock was identified by the content of energetic $He^{++}$ ions. Their results clearly indicate an increased wave (and in particular mirror mode) occurrence under quasi-perpendicular conditions. Génot et al. (2009) performed a statistical study of the occurrence of mirror structures over 5 years of Cluster observations using the GIPM (geocentric interplanetary medium) reference frame (Verigin et al., 2006), where fluctuations in the IMF direction are normalised away. Again, the results show a greater occurrence of mirror structures in the quasi-perpendicular hemisphere.

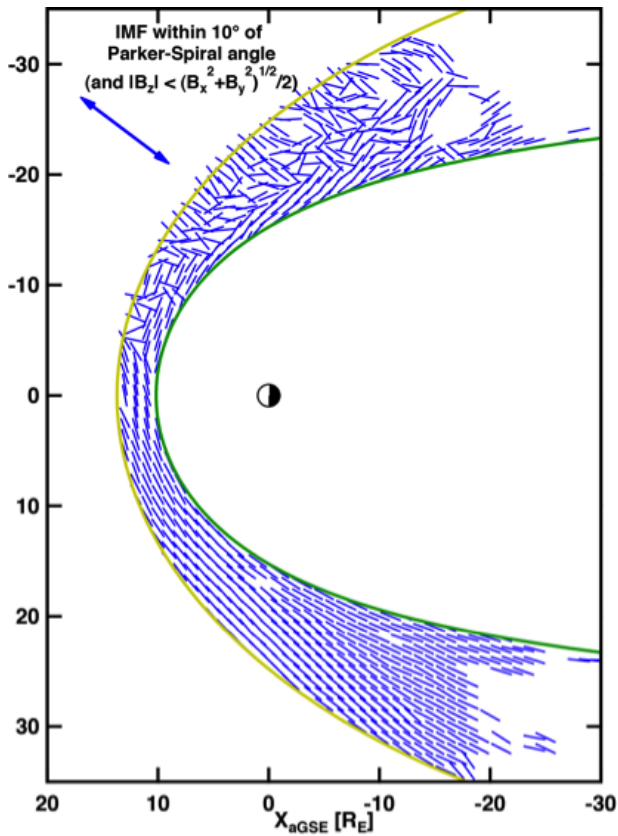

**Figure 4.** Median magnetic field vector orientation in $1 \times 1 \ R_E$ bins in the equatorial plane when IMF is within 10° of Parker spiral angle (adapted from Fig. 6 in Petrinec, 2013).

In summary, low-frequency field and plasma oscillations are ubiquitous in the magnetosheath and are organised according to upstream shock conditions. The quasi-parallel magnetosheath (found on the dawn side for predominant Parker spiral IMF) is typically more turbulent with large-amplitude and long wave period oscillations. On the other hand, quasi-perpendicular (predominantly dusk) magnetosheath oscillations are dominated by EMIC and mirror waves with smaller amplitudes and shorter wave periods. While this distinction is clearly observed in statistical studies and often in case studies, a large percentage of magnetosheath observations include a superposition of both effects (Fuselier et al., 1994). The identified asymmetries in observed field and plasma oscillations are summarised in Table 1.

## 2.2 Magnetopause asymmetries

The magnetopause is a thin current sheet separating the shocked magnetosheath plasma and its embedded interplanetary magnetic field on one side and the geomagnetic field on the other side. The current in the magnetopause is primarily caused by the differential motion of ions and electrons as they encounter the sharp magnetic gradient of the geomagnetic field. For a comprehensive overview of the magnetopause and its properties, we refer to, for example, Hasegawa (2012), so below we only focus on dawn–dusk asymmetries in the magnetopause.

Simultaneous measurements from both flanks of the magnetopause are rare. Also, the large variability in the thickness, orientation and motion of the magnetopause makes any direct comparison between the dawn and dusk flank magnetopause of little use. To our knowledge, the only study focusing explicitly on dawn–dusk asymmetries in macroscopic features of the magnetopause is the paper by Haaland and Gjerloev (2013). They used measurements from more than 5000 magnetopause traversals near the ecliptic plane by the Cluster constellation of satellites and reported significant and persistent dawn–dusk asymmetries in current density and magnetopause thickness.

Figure 5 shows the distribution of observed current densities for the dawn (red bars) and dusk (blue bars) magnetopause crossings during disturbed geomagnetic conditions. Most of the dawn magnetopause crossings have a current density around $10$–$15 \ \mathrm{nA \ m^{-2}}$, whereas the typical current density at dusk is around $25$–$30 \ \mathrm{nA \ m^{-2}}$. Mean current densities are 18 and $27 \ \mathrm{nA \ m^{-2}}$ for dawn and dusk, respectively. Haaland and Gjerloev (2013) noted that the dawn magnetopause was thicker, suggesting that the total current intensity on the two flanks were roughly equal. Two possible explanations for these dawn–dusk asymmetries are conceivable, both related to the boundary conditions. First, asymmetries in the magnetosheath as reported in Sect. 2.1.2 will influence the geometry and property of the magnetopause. A higher dusk-side magnetosheath magnetic field will cause a higher magnetic shear across the magnetopause, and thus a higher current density. Asymmetries in plasma parameters, in particular dynamic pressure, may also contribute, though simulations suggests that pressure enhancements are more likely to displace the magnetopause than compress it (Sonnerup et al., 2008). A second source of dawn–dusk asymmetry in magnetopause parameters are asymmetries in the ring current. In particular during disturbed conditions, the dusk sector of the ring current shows a faster energisation and higher current density than its dawn counterpart (Newell and Gjerloev, 2012). As a consequence, there will be a stronger magnetic perturbation at dusk and thus a higher magnetic shear across the magnetopause.

Several potential mechanisms by which plasma can enter the magnetosphere through the flank magnetopause have been suggested. These are thought to be most important when the magnetosphere is exposed to northward IMF, when the Dungey cycle (Dungey, 1961) does not dominate. These processes include transport via kinetic Alfvén waves (e.g. Johnson and Cheng, 1997), gradient drift entry (Olson and Pfitzer, 1985) and through rolled-up Kelvin–Helmholtz vortices (e.g. Fujimoto and Terasawa, 1994, 1995). Entry through double cusp (also known as dual lobe) type reconnection (Song and Russell, 1992) is also a possible mechanism during northward IMF. Asymmetries in reconnection at the dayside magnetopause under southward IMF, and the associated plasma entry, will be discussed in Sect. 3.1.

**Table 1.** Asymmetries in the average dayside magnetosheath.

| Process/property | Asymmetry preference | Source | Reference |
|---|---|---|---|
| Ion density | dawn, 19 % higher | Theory | Walters (1964) |
| | dawn, 33 % higher | IMP-8 (1978–1980) | Paularena et al. (2001) |
| | dawn, 1 % higher | IMP-8 (1994–1997) | Paularena et al. (2001) |
| | dawn, 21 % higher | THEMIS (2008–2010) | Walsh et al. (2012) |
| | dawn higher | Cluster (2001–2004) | Longmore et al. (2005) |
| $\mid B \mid$ | dusk, 23 % higher | THEMIS (2008–2010) | Walsh et al. (2012) |
| $T_i$ (ion temperature) | dawn, 33 % higher | Theory | Walters (1964) |
| | dawn, 12 % higher | THEMIS (2008–2010) | Walsh et al. (2012) |
| $\delta B$ (magnetic field jump) | dawn higher | IMP-4 (1967) | Fairfield and Ness (1970) |
| | higher at $Q_\parallel$ side | ISEE-2 (1977–1979) | Luhmann et al. (1986) |
| | higher at $Q_\parallel$ side | INTERBALL+Cluster (1996–2003) | Shevyrev et al. (2007) |
| Magnetic field turbulence | higher at $Q_\parallel$ side | Geotail (1996–2005) | Petrinec (2013) |
| Mirror mode occurrence | more frequent at $Q_\perp$ | Cluster (2001–2005) | Génot et al. (2009) |
| Kinetic wave occurrence | 91 % at $Q_\perp$, 40 % at $Q_\parallel$ | AMPTE CCE (1984) | Anderson and Fuselier (1993) |

Each of the mechanisms discussed above does not necessarily operate symmetrically with respect to the noon–midnight meridian, either because of their intrinsic properties or because of the dawn–dusk asymmetries in the magnetosheath as discussed in Sect. 2.1.2. This asymmetric plasma entry will also have consequences for the plasma sheet – see Sect. 2.3.2.

ULF waves in the magnetosheath can generate kinetic Alfvén waves (KAWs) when they interact with the magnetopause boundary (Johnson and Cheng, 1997) and in so doing stimulate the diffusive transport of ions into the magnetosphere. A recent survey by Yao et al. (2011) has shown that the wave power associated with KAWs is enhanced at the dawn magnetopause, which suggests enhanced transport on that flank. KAWs can heat ions both parallel (Hasegawa and Chen, 1975; Hasegawa and Mima, 1978) and, when they have a sufficiently large amplitude, perpendicular (Johnson and Cheng, 2001; Johnson et al., 2001) to the magnetic field, suggesting that if KAW-driven transport does preferentially occur on the dawn flank magnetopause it would also be associated with a heating of the transported magnetosheath plasma.

The growth of the Kelvin–Helmholtz instability may also have a dawn–dusk asymmetry. If finite Larmor radius effects are taken into account, growth is favoured on the duskside (Huba, 1996), while conditions in the magnetosheath under Parker spiral IMF conditions might favour growth on the dawn side (e.g. Engebretson et al., 1991). A statistical study of the occurrence of Kelvin–Helmholtz vortices on the flank magnetopause from Geotail data (Hasegawa et al., 2006) suggests no particular dawn–dusk asymmetry, although the majority of the detections were made antisunward of the terminator. An extension of this study by Taylor et al. (2012), including Double Star TC-1 data, did find an asymmetry with the occurrence of Kelvin–Helmholtz vortices favoured on the dusk flank magnetopause. However, this asymmetry was only present on the dayside. Simultaneous observations of Kelvin–Helmholtz vortices on both flanks are rare, and as such it is difficult to address any dawn–dusk asymmetry in their properties. However, Nishino et al. (2011) reported one observation of vortices occurring simultaneously on both flanks and showed that while their macroscopic properties were similar, on a microscopic level differences were observed, with more plasma mixing between magnetosheath and magnetospheric populations in the downside vortex than the duskside vortex.

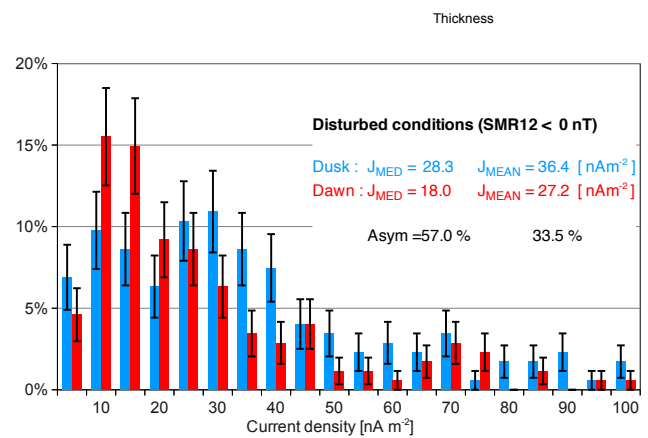

**Figure 5.** Distribution of magnetopause current densities based on Cluster curlometer results. Each bin in the histogram is $5\,\mathrm{nA\,m^{-2}}$ wide, and the indicated error bars are calculated as the square root of the number of observations in each bin and normalised. Red bars and values indicate dawn current densities, blue bars are corresponding dusk values. Mean, median and mode current density on dusk are significantly higher than their dawn counterparts. After Haaland and Gjerloev (2013).

Gradient drift entry naturally provides a dawn–dusk asymmetry: ions drift into the magnetosphere through the magnetopause on the dawn side, while electrons enter on the duskside (Olson and Pfitzer, 1985). However the efficiency of gradient drift entry and hence its potential to contribute to observed asymmetries in magnetospheric plasma is not well constrained. Treumann and Baumjohann (1988) calculated that only 5 % of magnetosheath particles that come into contact with the magnetopause become trapped, while through test particle simulations Richard et al. (1994) showed double cusp reconnection provided a much more efficient entry process. Indeed it is thought that double cusp reconnection operating under northward IMF is one of the dominant formation mechanisms for the cold dense plasma sheet (Lavraud et al., 2006). MHD simulations suggest that any dawn–dusk asymmetry in solar wind entry by double cusp reconnection is related to ionospheric conductance (Li et al., 2008a).

## 2.3 Magnetotail asymmetries

Throughout this review we will, in general, consider asymmetries about the noon–midnight meridian. Whilst at the boundaries of the magnetosphere such asymmetries are readily identifiable, as most of the boundaries are located well away from the meridian, within the magnetosphere asymmetries may depend on the coordinate system used. For example, the solar wind flow is not necessarily radial in the frame of the Earth; any non-radial flow will deflect the location of the central axis of the magnetosphere away from the $X_{GSM}$ axis (GSM = Geocentric Solar Magnetic – see e.g. Hapgood, 1997, for some commonly used coordinate systems and their definitions). The aberrated GSM (AGSM) coordinate system attempts to correct for this and has, for example, been shown to reduce the apparent asymmetry in convective flows in the magnetotail (Juusola et al., 2011).

### 2.3.1 Geometry and current systems

The magnetotail current sheet is often considered to be a static, Harris-type (Harris, 1962) current sheet separating the oppositely directed magnetic fields in the lobes. There is now sufficient evidence, particularly from the Cluster spacecraft, that the current sheet is in motion (e.g. Ness et al., 1967; Zhang et al., 2005; Sergeev et al., 2006; Forsyth et al., 2009), is bifurcated (Runov et al., 2006), or shows embedded current sheet signatures (Petrukovich et al., 2011) and is not, in fact, Harris-like in a statistical sense (Zhang et al., 2006; Rong et al., 2011). Statistical studies have also shown that the current sheet tends to be thinner, with a greater current density, on the duskward side of the magnetotail.

A number of multi-spacecraft analysis techniques have been developed to determine the current density within the current sheet and the sheet thickness (Dunlop et al., 1988; Shen et al., 2007; Artemyev et al., 2011). While the specifics of these techniques vary, they share a commonality that they all examine the currents based on magnetic field measurements by Cluster.

Statistically, the magnetotail current density measured by Cluster was consistently observed to be higher on the duskside than the dawn side of the magnetotail (e.g. Runov et al., 2005; Artemyev et al., 2011; Davey et al., 2012b). However, the values observed and the extent of the asymmetry between them differed for each study. On the duskside, the current densities ranged from 6 to 25 nA m$^{-2}$ (Artemyev et al., 2011) and on the dawn side, the current densities ranged from 4 to 10 nA m$^{-2}$. In contrast, the current sheet thickness was shown to be greater on the dawn side than on the duskside, both in absolute terms (Artemyev et al., 2011) and with respect to the local ion gyroradius (Rong et al., 2011). Rong et al. (2011) also showed that the probability of observing a thin current sheet was greater towards dusk. We note that the differences in current density and thickness tended to be comparable ($\sim$ 1.5–2.5 times difference), such that it appears that the total current flowing through the current sheet remains roughly constant.

It should be noted that the above studies by Runov et al. (2005), Artemyev et al. (2011), Rong et al. (2011) and Davey et al. (2012b) use different selection criteria to identify Cluster crossings of the tail current sheet. Rong et al. (2011) took any reversal of the $B_X$ component of the field to be a crossing, thus multiple small-scale fluctuations were identified as individual crossings, whereas Davey et al. (2012b) and Runov et al. (2005) required a change in $B_X$ between $\pm 5$ and $\pm 15$ nT respectively, with Runov et al. (2005) applying a further criterion that the duration of the field reversal was between 30 and 300 s. As such, Rong et al. (2011) identified 5992 crossings, Davey et al. (2012b) identified 279, and Runov et al. (2005) identified 78 events (although using only 1 year of Cluster data). Given the difference in the current sheet identifications and the number of events used in these studies, it is reassuring that the overall picture in their results is similar, even if the exact values differ. This difference may be a result of the different separations between the Cluster spacecraft throughout their lifetime (Runov et al., 2005; Forsyth et al., 2011).

Studies of the current sheet thickness and current density by Cluster rely on the phenomenon of "magnetotail flapping" (Speiser and Ness, 1967), whereby large-scale waves cause the current sheet to move locally in the $Z_{GSM}$ direction and to be tilted in the $YZ_{GSM}$ plane. The occurrence frequency of flapping increases towards dusk (Sergeev et al., 2006), but the tilt of the current sheet is greater towards dawn (Davey et al., 2012b). Furthermore, flapping has been shown to increase with substorm activity, but decrease with enhancements in the ring current (Davey et al., 2012a). Given that the thinning of current sheets during substorms is a well documented phenomenon (e.g. McPherron et al., 1973; Pulkkinen et al., 1994; Shen et al., 2008) one might expect thinner current sheets on average in the region in which most substorms occur (Frey et al., 2004; Frey and Mende, 2007). However, it is unclear from these results whether substorms are the cause or consequence of thin current sheets in this sector.

**Table 2.** Dawn–dusk asymmetries at the magnetopause and in plasma entry regions.

| Process/property | Asymmetry preference | Source | References |
|---|---|---|---|
| Magnetopause current density | dusk higher | Cluster (2001–2006) | Haaland and Gjerloev (2013) |
| Magnetopause thickness | dawn thicker | Cluster (2001–2006) | Haaland and Gjerloev (2013) |
| Kinetic Alfvén wave power | dawn larger | THEMIS | Yao et al. (2011) |
| Kelvin–Helmholtz wave growth | dawn larger | theory | Huba (1996) |
| | dawn larger | theory, ISEE, AMPTE | Engebretson et al. (1991) |
| Kelvin–Helmholtz vortex occurrence (dayside) | more at dawn | Double Star | Taylor et al. (2012) |
| Plasma mixing in Kelvin–Helmholtz vortices | more at dawn | Geotail, Cluster | Nishino et al. (2011) |

### 2.3.2 Nightside plasma sheet properties

Multiple ion populations exist in the magnetotail, including components with characteristic energies of 10s of eV (intense cold component), $\sim 300$–$600$ eV (cold component), $\sim 3$–$10$ keV (hot component), and $\sim 10$–$100$ keV (suprathermal). The higher ion density in the dawn flank magnetosheath leads to a higher density of cold component ions towards dawn in the magnetotail under northward IMF, as observed by C.-P. Wang et al. (2006). These ions have also been found to have higher temperatures at dawn than at dusk during northward IMF, in particular they are heated perpendicular to the magnetic field (Wing et al., 2005) and during intervals of high solar wind velocity (Wang et al., 2007). Nishino et al. (2007a) found the cold component ions to have parallel anisotropy ($T_{c\parallel} > T_{c\perp}$) at dusk, and conjectured that this is due to adiabatic heating during sunward convection. Wing et al. (2005) used Defense Meteorological Satellite Program (DMSP) satellites to infer plasma sheet temperatures and densities during periods of northward IMF. Their cold component density and temperature profiles are displayed in Fig. 6. The cold component density profile has peaks at dawn and dusk flanks, while the cold component temperatures are higher on the dawnside than the duskside, consistent with Hasegawa et al. (2003). This observation suggests that the magnetosheath ions have been heated in the entry process on the downside. The downside cold ion temperature is about 30–40 % higher than that on the duskside (see Fig. 6). Such asymmetric heating is consistent with the observed asymmetry in KAW transport described in Sect. 2.2.

In contrast, the hot component ions have higher temperatures toward dusk, especially within $\sim 20\,R_E$ of the Earth, due to the energy-dependent gradient–curvature drift. Spence and Kivelson (1993) developed a finite-width magnetotail model of the plasma sheet. In addition to a deep-tail source of particles, they found that including a particle source from the low-latitude boundary layer (LLBL) on the dawn side yields agreement with measurements of pressure and density. The model predicts a significant dawn–dusk asymmetry with higher ion pressure and temperature toward dusk for intervals of weak convection. Keesee et al. (2011) confirmed this model with average plasma sheet ion temperatures during quiet magnetospheric conditions calculated using energetic neutral atom (ENA) data from the TWINS mission, as seen in Fig. 7. This dawn–dusk asymmetry in ion temperatures has also been observed with in situ measurements by Geotail (Guild et al., 2008; C.-P. Wang et al., 2006). Using data from Geotail, Tsyganenko and Mukai (2003) derived a set of analytical models for the central plasma sheet density, temperature and pressure for ions with energies 7–42 keV in the $XY_{GSM}$ plane. Dawn–dusk asymmetries were found only within 10 $R_E$, near the boundary of their measurements, so were not included in their models that cover 10–50 $R_E$.

The contrasting ion temperature asymmetries between the hot and cold ion components during northward IMF yields measurements of two peaks in the ion distribution (the hot and cold components) on the dusk flanks, and one broad peak measured on the dawn flank (Fujimoto et al., 1998; Hasegawa et al., 2003; Wing et al., 2005). C.-P. Wang et al. (2006) measured the total ion density to be higher toward dawn for northward IMF, primarily due to the cold component ions, yielding equal pressures at dawn and dusk. They showed that the density asymmetry weakens during southward IMF, but the temperature asymmetry remains, yielding higher pressures at dusk. The magnetosphere $B_Z$ has been observed to be greater at dawn than at dusk (Fairfield, 1986; Guild et al., 2008; C.-P. Wang et al., 2006). This asymmetry serves to provide pressure balance to the higher densities at dusk. Both dawn and dusk flanks have high flux of ions with energies < 3 keV, with high flux extending toward the midnight meridian only from the dawn flank for intervals of northward IMF longer than an hour. This asymmetry

## Cold-component density and temperature profiles

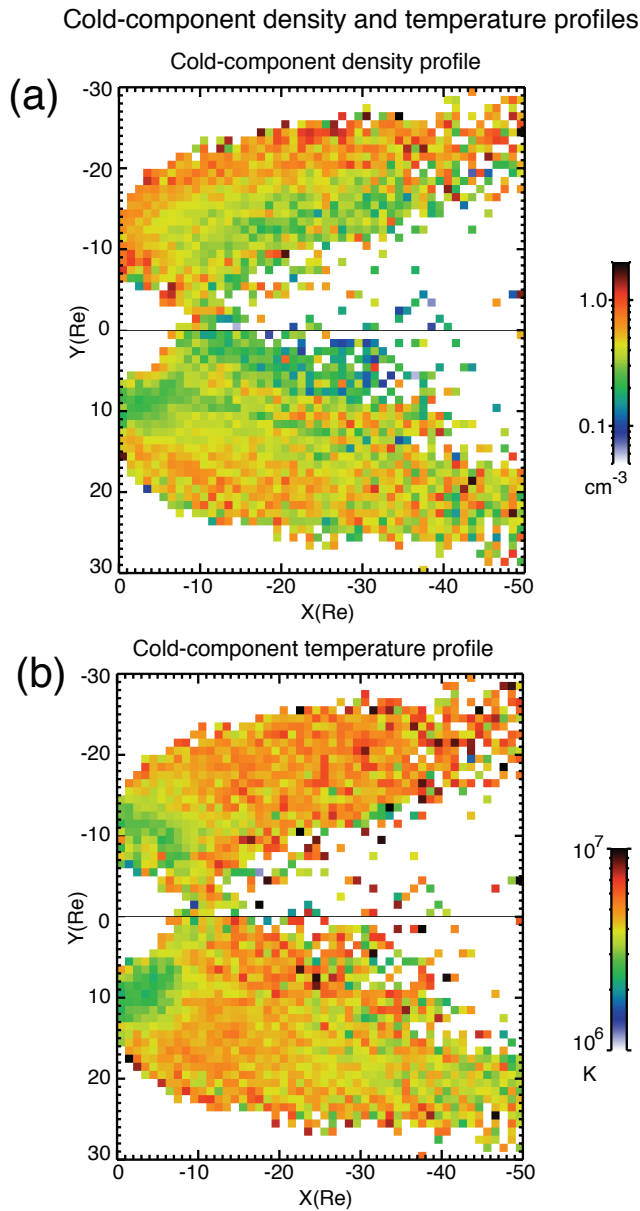

### Cold-component density profile

### (a)

### Cold-component temperature profile

### (b)

**Figure 6.** Density and temperature profiles of the cold component of the two-component Maxwellian distribution of the plasma sheet ions during northward IMF. Note the dawn–dusk asymmetry in the temperature profile, with the dawn flank ions having higher temperatures than the dusk flank ions. (from Wing et al., 2005).

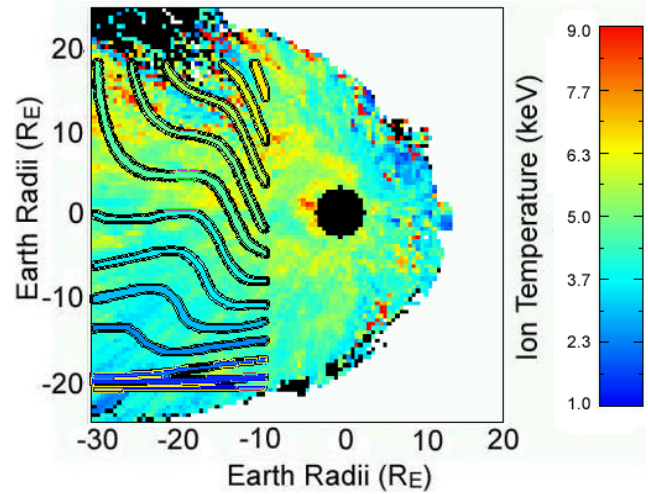

**Figure 7.** Ion temperatures calculated from TWINS ENA data mapped onto the $XY_{GSM}$ plane with the Sun to the right. A black disc with radius 3 $R_E$, centred at the Earth, indicates the region where analysis is not applicable. Contours of constant ion temperature as predicted by the finite tail width model of Spence and Kivelson (1993) are overlaid on the image. The measurements and model indicate higher plasma sheet hot component ion temperatures toward dusk during quiet magnetospheric conditions due to the gradient–curvature drift. (Adapted from Fig. 4 in Keesee et al., 2011).

is reduced during southward IMF as the high flux in the dawn sector decreases. For ions with energies > 6 keV, flux is higher at the dusk flank than the dawn flank, with the asymmetry being stronger for higher energies and southward IMF.

Both hot and cold components of the ions flow toward the midnight meridian under strong northward IMF conditions, due to (a) viscous interaction of the plasma sheet and the lobe and (b) vortical structures due to the Kelvin–Helmholtz instability (Nishino et al., 2007b). The average quiet time flow pattern in the plasma sheet displays a dawn–dusk asymmetry, with slower, sunward-directed flows post-midnight and faster, duskward-directed flows pre-midnight (Angelopoulos et al., 1993). The asymmetry in flow direction is also observed when averaging over all flow speeds (Hori et al., 2000), though the picture becomes somewhat more complicated when fast flows alone are examined (Sect. 2.4.2). The asymmetry in perpendicular flows is most significant within 10 $R_E$ of the midnight meridian (C.-P. Wang et al., 2006). The larger duskward component in the slow flow results from diamagnetic drift of ions due to the inward pressure gradient, which has a magnitude on the order of 25 km s$^{-1}$ (Angelopoulos et al., 1993).

Less is known about the intense cold component because ions in this energy range can only be detected when spacecraft are negatively charged as they pass through Earth's shadow. Seki et al. (2003) hypothesise that the intense cold component ions originate in the ionosphere because they have not undergone heating that would occur in the plasma sheet boundary layers. Similarly, measurements of the suprathermal component tend to be combined with the thermal component (Borovsky and Denton, 2010) or all components (Nagata et al., 2007), such that the specific dawn–dusk characteristics of this population have not been explored.

The electrons in the plasma sheet also exhibit a dawn–dusk asymmetry. Like the ions, there are two components of electrons, a hot component and a cold component (Wang et al., 2007; A. P. Walsh et al., 2013). Unlike the ions, however, both electron populations have been observed under northward and southward IMF, although a two-component electron plasma sheet is more likely to be observed under southward IMF (A. P. Walsh et al., 2013). Under southward IMF the two-component electron plasma sheet is more likely to be observed in the pre-midnight sector than the post-midnight sector. Under northward IMF the occurrence follows the pattern of the large-scale Birkeland currents coupling the ionosphere and magnetosphere – a two-component electron plasma sheet is more likely to be observed mapping to lower latitudes in the pre-midnight sector and higher latitudes in the post-midnight sector. This suggests the cold electrons have their source in the ionosphere, rather than the solar wind, and are transported to the plasma sheet via downward field-aligned currents (Iijima and Potemra, 1978; A. P. Walsh et al., 2013).

## 2.4 Asymmetries in magnetotail dynamics

### 2.4.1 Substorms and other modes

Southward-pointing IMF results in a circulation of magnetic flux in the magnetosphere – with dayside reconnection opening flux, transportation of open flux into the lobes, nightside reconnection closing flux to form the plasma sheet, and return of flux back to the dayside (Dungey, 1961). The magnetosphere is driven to many modes of response due to magnetic reconnection with the solar wind IMF. These include substorms, magnetic storms, steady magnetospheric convection, and sawtooth events, as well as smaller responses such as pseudobreakups and poleward boundary intensifications (for a full review of these modes, see e.g McPherron et al., 2008). These events with enhanced sunward convection in the plasma sheet will dominate over certain asymmetries discussed above, such as the quiet-time dawn–dusk thermal pressure asymmetry (Spence and Kivelson, 1990).

The most common and well-studied mode of response is the substorm. Numerous researchers have found asymmetries in the average substorm onset location, with the most likely onset shifted duskward to 23:00 MLT (Frey and Mende, 2007, and references therein). The onset MLT of substorms is strongly influenced by the IMF clock angle, which shifts the dayside reconnection geometry in such a way as to create a "tilted" configuration away from direct noon–midnight reconnection (Østgaard et al., 2011). Internal factors, such as solar illumination and its effects on ionospheric conductivity, can also influence the average onset location in latitude and local time (Wang et al., 2005, see also Sect. 3.2). Sawtooth events also display dawn–dusk asymmetry, with intense tail reconnection signatures occurring premidnight (Brambles et al., 2011). The sawtooth asymmetry is attributed to ion outflow asymmetry which is in turn a result of ionospheric conductance asymmetry. Many dynamic signatures of enhanced convection, especially during substorms, also display a pre-midnight occurrence peak. These include magnetic reconnection, bursty bulk flows, transient dipolarisations and energetic particle bursts and injections, described in more detail below.

Recently Nagai et al. (2013) surveyed a large data set including Geotail observations from 1996 to 2012 in the area of $-32 < X_{AGSM} < -18$ $R_E$ and $|Y_{AGSM}| < 20$ $R_E$. Active reconnection events were selected using the following criteria: (1) $|B_X| < 10$ nT to select plasma sheet samples, (2) $V_{i_X} < -500$ km s$^{-1}$ and $B_Z < 0$ to select tailward fast flows, (3) earthward flow at $V_{i_X} > 300$ km s$^{-1}$ and $B_Z > 0$ observed within 10 min after the tailward flow to select the flow reversals, and (4) $V_{eY} < -1000$ km s$^{-1}$ during at least one sample within 48 s long interval around the flow reversal instant to select the active reconnection when electrons undergo substantial acceleration; 30 active reconnection events were selected. The analysis of occurrence rate distribution has shown that events may be found in the sector $-6 < Y_{AGSM} < 8$ $R_E$. The occurrence rate is considerably higher in the pre-midnight sector $0 < Y_{AGSM} < 8$ $R_E$.

Slavin et al. (2005) used Cluster observations to study travelling compression regions (TCRs), which are commonly accepted to be remote signatures of a reconnection outflow in the magnetotail lobes at distances $-19 < X < -11$ $R_E$, and noticed a dawn–dusk asymmetry in the event distribution in the $XY_{AGSM}$ plane with considerably larger number of events observed in the pre-midnight sector. Similarly, Imber et al. (2011) inferred the dawn–dusk location of the reconnection site from statistical studies of THEMIS observations of flux ropes and TCRs during the time period December 2008 to April 2009. Magnetic signatures, including a bipolar variation in $B_Z$ passing through $B_Z = 0$ and an enhancement in $B_Y$ at $B_Z = 0$ were used to identify a flux rope. A bipolar $\Delta B_Z$ signature relative to the background field and total field variation with $(\Delta B)/B > 1$ % were used to identify TCRs; 87 events (both flux ropes and TCRs) were identified. Plotting the spacecraft location for all the events in the $XY_{AGSM}$ plane, Imber et al. (2011) have shown an obvious dawn–dusk asymmetry with 81 % of events observed in the dusk sector. The event probability (number of events per unit time) also showed strong duskward asymmetry: a peak of the Gaussian fit to the data is at $Y_{AGSM} = 7.0$ $R_E$ and the full width at half maximum is 15.5 $R_E$.

In their survey of magnetotail current sheet crossings, Rong et al. (2011) found that 329 out of 5992 current sheet crossings by the Cluster spacecraft in 2001, 2003 and 2004 had a negative $B_Z$ component. These negative $B_Z$ current sheet crossings were predominantly found to occur at azimuths of 110° to 210° and had field curvature directions pointing away from the Earth. Given that $B_Z$ is expected to be positive on closed magnetic field lines in the magnetotail plasma sheet, Rong et al. (2011) interpreted these observations as showing that reconnection was "more inclined to be triggered in current sheet regions with MLT being

**Table 3.** Dawn–dusk asymmetries in magnetotail processes and properties.

| Process/property | Asymmetry preference | Source | Reference |
|---|---|---|---|
| Hot component of ion temperature | higher at dusk | Geotail | Guild et al. (2008); C.-P. Wang et al. (2006) |
| | higher at dusk | TWINS | Keesee et al. (2011) |
| | higher at dusk | model | Spence and Kivelson (1990) |
| Cold component of ion temperature | higher at dawn | DMSP | Wing et al. (2005) |
| Cold component of ion density | higher at dawn | DMSP | Wing et al. (2005) |
| $B_Z$ | higher at dawn | IMP 6, 7, 8 | Fairfield (1986) |
| | | Geotail | Guild et al. (2008); C.-P. Wang et al. (2006) |
| Flankward quiet time flows | more frequent in pre-midnight | ISEE 2 | Angelopoulos et al. (1993) |
| | more frequent in pre-midnight | Geotail | Hori et al. (2000) |

∼ 21:00–01:00", thus showing a clear dawn–dusk asymmetry in the distance downtail at which reconnection occurs.

Reconnection signatures observed in the distant tail and at lunar orbit also exhibit dawn–dusk asymmetry. Slavin et al. (1985) have studied average and substorm conditions in the distant magnetotail using ISEE-3 data. It was found that negative $B_Z$ and fast tailward flow was predominantly observed in the pre-midnight sector ($0 < Y_{GSM} < 10\,R_E$ at $-100 > X > -180\,R_E$). Further tailward, at $-180 > X > -120\,R_E$, the region of predominant $B_Z < 0$ and fast tailward flow expands azimuthally to a broad region between $Y_{GSM} = 0$ and $\sim -20\,R_E$. It should be noted, though, that at those geocentric distances the GSM coordinate system may not be appropriate, and the broad distribution of $-B_Z$ and $-V_X$ maxima may be an apparent effect of averaging over different solar wind/IMF conditions.

Recently reconnection outflows and plasmoid observations by two ARTEMIS spacecraft in lunar orbit have been statistically studied (Li et al., 2014). That study revealed a dawn–dusk asymmetry with occurrence rate of plasmoid observations higher within $-2 < Y_{AGSM} < 12\,R_E$. The occurrence distribution has a similar but broader pattern compared with previous studies on plasmoids or reconnection flow reversals in the near-Earth region (Imber et al., 2011; Nagai et al., 2013).

### 2.4.2 Fast flows in the plasma sheet

Fast plasma flows in the magnetotail above a "background" convection velocity are often associated with substorm activity as a key device by which closed magnetic flux can be transported towards the inner magnetosphere and as a possible mechanism for the triggering of instabilities in the inner magnetosphere that lead to substorm onset (Baumjohann et al., 1990). Short (sub-minute) bursts of enhanced plasma flow (termed flow bursts) are most likely generated by impulsive magnetotail reconnection (see Sect. 2.4.1). The flow bursts are grouped into ∼ 10 min events known as bursty bulk flows (BBFs) (Angelopoulos et al., 1992), although these terms are sometimes used interchangeably throughout the literature. Numerous statistical studies of BBFs, conducted during last the two decades, result in rather controversial conclusions on asymmetries in the azimuthal (MLT) dependence of BBF distribution. Comparison between them is complicated by the use of different selection criteria to identify individual events.

A set of studies applying selection criteria based upon either magnetic field ($(B_X^2 + B_Y^2)^{1/2} < 15\,nT$) or $\beta > 0.5$ to select plasma sheet samples and flow velocity magnitude ($|V_X| > 400\,km\,s^{-1}$) to select flow bursts (FB) and BBF events did not reveal a pronounced dawn–dusk anisotropy in the event distribution (Baumjohann et al., 1990; Angelopoulos et al., 1994). Some asymmetry in velocity magnitudes with faster flows observed in the pre-midnight sector were considered apparent and attributed to orbital biases (Nakamura et al., 1991). On the other hand, studies of Geotail, WIND and THEMIS data with selection criteria differentiating convective flows (i.e. perpendicular to the instantaneous magnetic field) and field-aligned beams resulted in pronounced asymmetry in the convective flow distributions and symmetric field-aligned beam distributions (Nagai et al., 1998; Raj et al., 2002; McPherron et al., 2011).

Statistical analysis of plasma bulk velocity observed by Cluster during neutral sheet ($|B_X| < 5\,nT$) crossings at radial distances $R \approx 18\,R_E$ revealed dawn–dusk asymmetries in the horizontal velocity magnitude ($V_{eq} = (V_X^2 + V_Y^2)^{1/2}$) with larger values ($V_{eq} > 400\,km\,s^{-1}$) in the pre-midnight sector of the magnetotail within $0 < Y_{AGSM} < 10\,R_E$. The average equatorial velocity in the post-midnight sector did not exceed $200\,km\,s^{-1}$ (Runov et al., 2005). Conversely, a study of the comprehensive data set that includes 15 years of Geotail, Cluster and THEMIS observations in the magnetotail applying the criterion $\beta > 0.5$ to select plasma sheet samples revealed no asymmetry tailward of $X = -15\,R_E$ in the aberrated coordinate system (Juusola et al., 2011). Closer to Earth, the average convection at a velocity smaller than $200\,km\,s^{-1}$ shows some duskward asymmetry. This asymmetry was attributed to the ion gradient drift close to the inner edge of the plasma sheet (see also Hori et al., 2000). The distribution of higher velocity remains fairly symmetric with respect to the midnight in AGSM (Juusola et al., 2011).

The dawn–dusk asymmetry in the magnetotail plasma flows also depends on the level and character of geomagnetic activity. Recent studies of Geotail and THEMIS observations over a span of 14 years comparing the convection patterns observed during periods of steady magnetospheric convection (SMC) and substorm phases have revealed that the probability of earthward fast flows ($V_{XY} > 200\,\mathrm{km\,s^{-1}}$) is fairly symmetric with respect to midnight for SMC but slightly asymmetric with a peak at $\sim 23{:}00$ MLT during substorm growth phases. This duskward asymmetry vanishes during expansion and recovery substorm phases (Kissinger et al., 2012).

To summarise, the statistical studies of BBFs and plasma convection in the magnetotail conducted so far do not provide any definitive answer on the question on dawn–dusk asymmetry in the flow pattern. The results strongly depend on the selection criteria. More specifically, studies with criteria based upon the perpendicular velocity tend to show the duskward asymmetry. Conversely, the studies based upon $|B_{XY}|$ and $\beta$-related criteria typically result in a fairly symmetric flow pattern. Another important issue is the selection of fast flow events and differentiation of them from the background convection. It was noticed in observations that BBFs (flow bursts) are typically associated with (1) increased northward (southward) magnetic field component ($B_Z$) and (2) decrease in the plasma density (Angelopoulos et al., 1992, 1994; Ohtani et al., 2004). These characteristics, attributed to so-called "plasma bubbles" (e.g. Chen and Wolf, 1993; Wolf et al., 2009), may be used to differentiate transient BBFs from the steady convection. The rapid increase in $B_Z$ and simultaneous decrease in the plasma density were recently found to be characteristics of dipolarisation fronts (Runov et al., 2011; Liu et al., 2013) that will be discussed in the next section.

### 2.4.3 Transient dipolarisations and dipolarisation fronts

Russell and McPherron (1973) first reported observations of front-like, spatially and temporally localised, sharp increases in the northward magnetic field component $B_Z$. Timing of the two-point observations by OGO-5 (at $X = -8.2\,R_E$) and ATS-1 (at $X = -5.6\,R_E$) spacecraft indicated earthward propagation of this magnetic structure. Later it was found that the $B_Z$ enhancement is accompanied by BBFs (Angelopoulos et al., 1992; Ohtani et al., 2004). The enhanced $V \times \boldsymbol{B}$-electric field (magnetic flux transfer rate) appeared in the form of $\sim 100\,\mathrm{s}$ long pulses, referred to as rapid flux transfer events (Schödel et al., 2001). For such structures, the $B_Z$ enhancements are spatial structures travelling with the flow.

At other times, particularly in the inner magnetosphere, plasma flows are not observed during the $B_Z$ enhancements; in these cases the $B_Z$ enhancements do not contribute to local flux transport and are the result of non-local currents from a substorm current wedge, (e.g. McPherron et al., 1973) most often tailward of the observation point (a remote-sensing effect – see, e.g. Nagai, 1982). Both types of events have been intensely studied in the past under various names, such as nightside flux transfer events (e.g. Sergeev et al., 1992), flux pileup (Hesse and Birn, 1991; Shiokawa et al., 1997; Baumjohann et al., 1999) and current disruption (e.g. Lui, 1996). Treated as flowing spatial structures, the sharp $B_Z$ enhancements have been referred to as "dipolarisation fronts" (e.g. Nakamura et al., 2002; Runov et al., 2009).

It has been shown that the earthward-propagating dipolarisation fronts are associated with a rapid decrease in the plasma density and embedded into the earthward plasma flow (Runov et al., 2009, 2011). The fronts are thin boundaries (with the thickness of an ion thermal gyroradius), separating underpopulated dipolarised flux tubes, often referred to as "plasma bubbles" (e.g. Wolf et al., 2009), and the ambient plasma sheet population. Most likely, the dipolarisation fronts are generated in the course of impulsive magnetic reconnection in the mid or near magnetotail (see e.g. Runov et al., 2012, and references therein). Alternatively, the fronts may appear as a result of kinetic interchange instability in the near-Earth plasma sheet (Pritchett and Coroniti, 2010).

Recently, Liu et al. (2013) statistically studied several hundred dipolarisation fronts observed by THEMIS probes in the plasma sheet at $-25 < X < -7\,R_E$ and at variety of azimuthal ($Y$) positions. The events were selected using a set of selection criteria based mainly on magnetic field and rate of magnetic field changes. The selected events may, therefore, include those of all categories discussed above. The analysis has shown, however, that the increase in $B_Z$ was associated with the rapid decrease in plasma density and was embedded into earthward plasma flow. Thus, the majority of selected events were dipolarisation fronts. Figure 8 shows (a) the distribution of selected events and (b) the occurrence rate of the dipolarisation fronts in the $XY_{\mathrm{GSM}}$ plane. The event distribution shows a pronounced dawn–dusk asymmetry with more events observed in pre-midnight sector within $0 < Y < 8\,R_E$. The occurrence rate exhibits a maximum in $2 < Y < 6\,R_E$ bins in a range of $-20 < X < -7\,R_E$.

Dipolarisation fronts are typically embedded into fast earthward flows (BBFs). However, as was shown in the previous section, contrary to that of the dipolarisation fronts, azimuthal distribution of BBF occurrence rate does not display any pronounced dawn–dusk asymmetry. Nonetheless, because of large $B_Z$, the magnetic flux is transported mainly by the dipolarisation fronts (Liu et al., 2013). Thus, the magnetic flux transport is strongly asymmetric with respect to the midnight meridian with maximum of the occurrence rate distribution between $0 < Y < 8\,R_E$. This sector of the magnetotail is also the area of maximum probability of magnetotail reconnection (see Sect. 2.4.1).

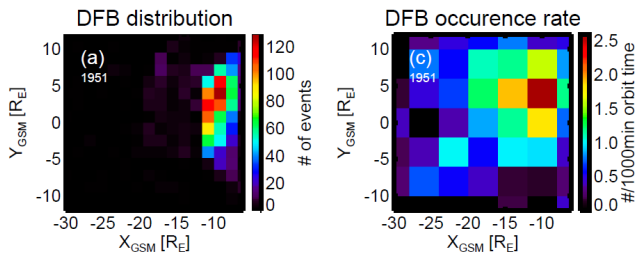

**Figure 8.** Statistical distribution and occurrence rate of dipolarisation fronts observed by THEMIS during the 2007–2011 tail seasons. After Liu et al. (2013).

### 2.4.4 Energetic particle injections

Observations of energetic particles at geosynchronous orbit (GEO) revealed sudden increases in the particle fluxes that are typically observed during enhanced geomagnetic activity (substorms and storms) and referred to as "energetic particle injections" (e.g. McIlwain, 1974; Mauk and Meng, 1987; Birn et al., 1997a, 1998). The injections observed at GEO fall into two distinct categories: dispersionless and dispersed. In the former case, the enhancement in particle fluxes at different energies occurs roughly simultaneously, whereas in the latter case a pronounced delay between the flux enhancement at different energies is observed (see e.g. Birn et al., 1997a). A commonly accepted explanation for these two types of injections is that dispersionless injections are observed by a satellite situated in or near the source of accelerated particles, whereas dispersed injections are observed by a satellite that is azimuthally distant from the injection source region, so that gradient and curvature drifts are responsible for the delay in arrival times of particles of different energies (e.g. Anderson and Takahashi, 2000; Zaharia et al., 2000).

A pronounced dawn–dusk asymmetry has been found in spatial distributions of ion and electron injection observed at GEO. It has been found that local time (LT) distribution of the occurrence frequency of high-energy ($> 2$ MeV) electron flux increase events is asymmetric with respect to midnight with a larger rate in the dusk sector (Nagai, 1982). The dawn–dusk asymmetry in the MeV electron fluxes was explained by an increase in ion pressure in the duskside inner magnetosphere during enhanced convection that leads to a magnetic field decrease due to diamagnetic effect and, therefore, to the adiabatic decrease in electron flux. Lopez et al. (1990) studied dispersionless ion injections observed by AMPTE as a function of local time and radial distance. They found an occurrence peak near midnight, with asymmetry towards pre-midnight local times. A similar study, but using electron injection measurements from the CRRES satellite was conducted by Friedel et al. (1996) Their analysis showed that the region of dispersionless injections is sharply bounded in magnetic local time and can have a radial extent of several $R_E$.

Birn et al. (1997a) studied properties of the dispersionless injections observed at GEO by Los Alamos 1989-046 satellite, situated near the magnetic equator in the midnight sector of the magnetotail. Their analysis revealed a significant asymmetry in the injection properties with respect to the Magnetic Local Time (MLT): proton-only injections are predominantly observed in the evening and pre-midnight sectors (18:00–00:00 MLT), whereas electron-only injections are observed in the post-midnight sector (00:00–05:00 MLT). Near midnight, the probability of both ion and electron injection observations maximises. Another finding is that the probability to observe first proton then electron injections maximises between 21:00 and 23:00 MLT, whereas the probability to observe first electron then proton injections is larger at midnight and in the post-midnight sector (23:00–03:00 MLT).

The azimuthal offset of ion and electron dispersionless injections was confirmed by the simultaneous observations by two closely spaced synchronous satellites (Thomsen et al., 2001). Similar results were also obtained by Sergeev et al. (2013), who compared MLT distributions of proton and electron dispersionless injections and auroral streamers. It was shown that proton (electron) injections are seen exclusively at negative (positive) $\Delta$ MLT, where $\Delta$ MLT is the difference between MLTs of injection and streamer observations (MLT$_{sc}$–MLT$_{str}$). Test particle tracing in magnetic and electric fields resulting from MHD simulations of magnetotail reconnections also showed that ion and electron dispersionless injection boundaries spread azimuthally duskward and dawnward, respectively (Birn et al., 1997b; Birn et al., 1998).

It is important to emphasise that *dispersionless* injections were studied in the above discussed works. Thus, the spatial dawn–dusk asymmetry in ion and electron injections cannot be attributed to the gradient and curvature drifts in the background quasi-dipole field that will lead the energy dispersion. Recent studies, both observation- and test-particle-simulation-based, have revealed that the dawn–dusk asymmetry appears within the fast-flow channel, where $B_Z$ is larger than in the surrounding plasma sheet and, therefore, in the steady-state reference frame, the electric field (mainly $V \times B$) is enhanced (Birn et al., 2012; Gabrielse et al., 2012; Runov et al., 2013). Although this asymmetry is due to ion (electron) duskward (dawnward) drift within the channel, because of finite channel cross-tail size (1–3 $R_E$, Nakamura et al., 2004) it does not lead the significant energy dispersion.

Injections have also been observed in the outer magnetotail. Bursts of high-energy protons and electrons with durations varying from 100 s to 100s of minutes were observed by IMP-7 at geocentric distance $\sim 35\,R_E$ (e.g. Sarris et al., 1976). Proton bursts were observed equally frequently in the dawn- and dusksides of the magnetotail. However, a strong dawn–dusk asymmetry in the distribution of the intense proton bursts $> 500$ (cm$^2$ s sr MeV)$^{-1}$ with majority of these occurring in the dusk magnetotail was revealed. To our knowledge, no dawn–dusk asymmetry in high-energy electron bursts has been found in the outer magnetotail.

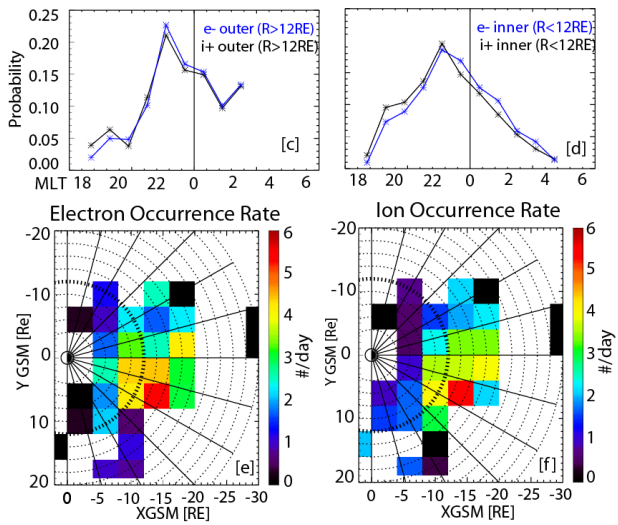

**Figure 9.** Probability and occurrence rate of ion (black) and electron (blue) dispersionless injections observed by THEMIS. From Gabrielse et al. (2014).

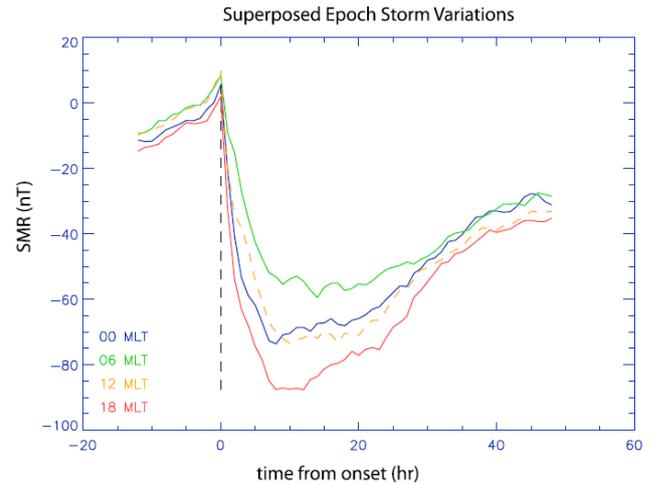

**Figure 10.** The four regional SMR indices from a superposed epoch study of 125 storms. (Adapted from Fig. 7 in Newell and Gjerloev, 2012).

THEMIS observations of ion and electron dispersionless injections at geocentric distances from 6 to $\sim 20\,R_{\mathrm{E}}$ were recently statistically studied by Gabrielse et al. (2014). That study demonstrated (see Fig. 9) that the injections observed far beyond geosynchronous orbit exhibit a pronounced dawn–dusk asymmetry. Specifically, (1) at all distances both ion and electron injections are more frequently observed in the pre-midnight sector with a peak in probability at $\sim 23{:}00$ MLT, (2) at radial distances larger that $12\,R_{\mathrm{E}}$ (outer region) the probability to detect ion and electron injections is quite similar with the electron injection probability offset slightly downward of the 23:00 MLT peak, (3) within $12\,R_{\mathrm{E}}$ (inner region) the probability distributions for both i$^{+}$ and e$^{-}$ injections are broader than that in the outer region; the electron injection probability being shifted notably towards dawn from the 23:00 MLT peak.

### 2.4.5 Magnetotail asymmetries – summary

Numerous observations suggest that dynamic processes in the magnetotail occur predominantly on the duskside and, typically, localised within several $R_{\mathrm{E}}$ in the pre-midnight sector (Table 4). The localisation of convective fast flows, dipolarisation fronts and dispersionless particle injections, plasmoids and TCRs can be understood by considering these events as direct or indirect consequences of magnetic field energy release via magnetotail reconnection. Reconnection, in turn, is more probable within the pre-midnight sector because the cross-tail current density is higher and the current sheet is thinner. What determines the reduced current sheet thickness in the pre-midnight sector remains an open question.

### 2.5 Inner magnetosphere asymmetries

The inner magnetosphere is the region of the magnetosphere closest to the Earth, reaching out from the ionosphere to the magnetopause on the dayside and to $\sim 8$–$10\,R_{\mathrm{E}}$ on the nightside (exclusive of the polar regions). The structure and dynamics of the inner magnetosphere are driven by input from the ionosphere and magnetotail and the interaction of this material with the dipole magnetic field lines. Energetic particles are trapped in this region and undergo a variety of drift motions due to the gradient and curvature of the magnetic field (e.g. Schulz and Lanzerotti, 1974), with electrons drifting eastward/dawnward and ions westward/duskward. We detail asymmetries that occur in the radiation belts, ring current, and plasmasphere regions. Many are likely the result of a zoo of wave–particle interactions, which are discussed separately.

### 2.5.1 Ring current symmetries

Dusk–dawn asymmetries in the ring current have been known since 1918 when Chapman (1918) observed a more pronounced disturbance in the north–south ($H$) component of Earth's magnetic field at dusk. The stronger storm-time disturbance at dusk is generally attributed to the partial ring current (Harel et al., 1981). Love and Gannon (2009) found the difference between the dusk and dawn disturbance to be linearly proportional to the Dst index. Tsyganenko et al. (2003) modelled the storm-time disturbance of Earth's magnetic field using satellite-based magnetometer data for events with Dst minimum at least $-65\,\mathrm{nT}$ and found a stronger disturbance at dusk. Newell and Gjerloev (2012) introduced the SMR (SuperMag Ring current) indices that indicate the average perturbation of the horizontal component of the Earth's magnetic field measured by a set of ground magnetometer stations centred at four local times: SMR-00, SMR-06, SMR-12 and SMR-18. In a superposed epoch analysis of 125 storms, they found a consistently stronger perturbation at dusk, as seen in Fig. 10. Using an enhanced TS04 model, Shi et al. (2008) modelled the perturbation in the $H$ of the low- to mid-latitude geomagnetic field to determine the contributions of various currents, including the region 1 and 2 field-aligned currents, currents that close the Chapman–Ferraro current in the magnetopause and through the partial ring current, respectively. For a weak partial ring current, they found a day–night asymmetry with negative $H$ perturbation around noon and positive $H$ perturbation around midnight, primarily caused by region 1 field-aligned currents. During storm main phase, the partial ring current tended to be stronger, pushing the negative $H$ perturbations toward dusk, yielding a dawn–dusk asymmetry. Solar wind dynamic pressure enhancements tend to increase the partial ring current and field-aligned currents, resulting in nearly instantaneous measurements of the dawn–dusk asymmetry in $H$ perturbations. The strength of the partial ring current during a storm depends on preconditioning based on northward or southward IMF $B_Z$.

**Table 4.** Asymmetries in the magnetotail dynamics.

| Process | Asymmetry preference | Source (years) | Reference |
|---|---|---|---|
| Near-tail reconnection signatures | more frequent at dusk | Geotail, Cluster (1996–1912) | Eastwood et al. (2010) |
| | more frequent at dusk | THEMIS (1996–2012) | Nagai et al. (2013), Imber et al. (2011) |
| Plasmoid/TCR in the mid and distant tail | more frequent at dusk | ISEE-3 (1982–1983) | Slavin et al. (1985) |
| | more frequent at dusk | ARTEMIS (2010–2012) | Li et al. (2014) |
| Bursty bulk flows | ambiguous | AMPTE, ISEE-1/2 | Baumjohann et al. (1990), Angelopoulos et al. (1992) |
| | | Geotail, Cluster,THEMIS | Juusola et al. (2011) |
| Convective flows | more frequent at dusk | Geotail, WIND, THEMIS | Nagai et al. (1998), Raj et al. (2002) |
| | | | McPherron et al. (2011) |
| Dipolarisation fronts | more frequent at dusk | THEMIS (2007–2012) | Liu et al. (2013) |
| Particle injections | more frequent at dusk | GEO | Birn et al. (1997a) |

Using simulations, Ebihara and Ejiri (2003) explained that the asymmetry in the magnetic field causes protons with small pitch angles to drift toward earlier local times than protons with larger pitch angles. Ring current ions move along equipotential surfaces while the first and second adiabatic invariants are conserved, leading to adiabatic heating toward dusk and cooling toward dawn (Mililllo et al., 1996). Skewed equatorial electric fields produced by the closure of the partial ring current during active periods cause the peak in the proton distribution function to occur between midnight and dawn, as observed in ENA images such as Fig. 11.

### 2.5.2 Radiation belt asymmetries

Dawn–dusk asymmetries in radiation belt particle fluxes are not well studied; instead much research has focused on the source and loss processes that do preferentially act at certain local times (see recent reviews by Millan and Thorne, 2007; Thorne, 2010, for example). Many of these source and loss processes are related to wave–particle interactions and hence occur in the regions to be described in Sect. 2.5.4. Changes in radiation belt particle fluxes can also be observed, not as a result of particle acceleration or loss to the atmosphere, but instead through the displacement of the drift shells on which the particles travel. This displacement is dependent on the geometry of the magnetic field in the inner magnetosphere and hence on the strength of the ring current – the so-called Dst effect (McIlwain, 1966; Williams et al., 1968). Thus, any asymmetries in ring current strength can alter the drift paths of radiation belt electrons which manifests as an asymmetry in electron flux. There is also evidence for a dawn–dusk asymmetry in radiation belt electron flux caused by substorm-related changes in the inner magnetospheric magnetic field: a more tail-like magnetic field in the dusk sector shifts the drift path of energetic electrons, effectively moving the radiation belt to lower latitudes (Lazutin, 2012).

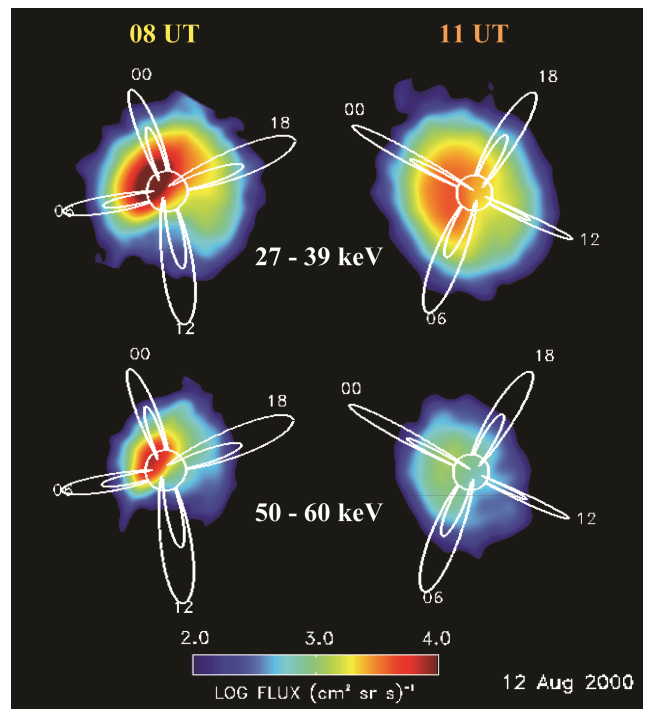

**Figure 11.** Images from two energy channels, 27–39 keV (top row) and 50–60 keV (bottom row), from the High Energy Neutral Atom (HENA) instrument on the IMAGE mission at two times during the 12 August 2000 geomagnetic storm, 08:00 UT (just before minimum Dst, left column) and 11:00 UT (just after minimum Dst, right column). The limb of the Earth and dipole field lines ($L = 4$ and $L = 8$) at 00:00, 06:00, 12:00 and 18:00 MLT are shown in white. The proton distribution peak occurs in the midnight–dawn sector due to skewed equatorial electric fields produced by the closure of the partial ring current during active periods. (Adapted from Fig. 7 in Fok et al., 2003, .)

### 2.5.3 Plasmasphere asymmetries

The upward extension of the cold, dense plasma from the Earth's ionosphere forms the plasmasphere. Motion of the plasmaspheric population is governed by an electric field made up of two potential components, corotation and convection. The first potential dominates close to the Earth and is an effect of Earth's own rotation. The second comes from the coupling of the solar wind and the magnetosphere and is a result of sunward return of plasma sheet flow. Figure 12 shows how cold particles drift under such potentials. During geomagnetically quiet times, the plasmaspheric particles travel on closed $E \times B$ drift shells around the Earth (within the separatrix), maintaining a fairly steady population. During disturbed times, when dayside reconnection increases, the convection potential is enhanced. An increase in the convection potential will cause an inward motion of the edge of the plasmasphere, or the plasmapause, and erosion of the outer material (Grebowsky, 1970; Chen and Wolf, 1972; Carpenter et al., 1993). Erosion of the outer plasma forms a sunward convecting drainage plume or the plasmaspheric plume.

Recent spacecraft measurements with Cluster and THEMIS as well as imaging from IMAGE have provided insight to the morphology of plumes. During storm onset the dayside plasmasphere surges sunward over a wide extent in local time. As time progresses during the disturbance, the extension narrows on the dawn side while staying relatively stationary in the dusk extension (Sandel et al., 2001; Goldstein et al., 2005). When dayside reconnection decreases the narrow plume typically rotates eastward and wraps itself around the plasmasphere (Goldstein et al., 2004; Spasojević et al., 2004).

The extension of cold dense plasma from the plume transports a large amount of mass to the outer magnetosphere. Borovsky and Denton (2008) estimates that $2 \times 10^{31}$ ions (34 tonnes of protons) are transported via plumes in the life of a storm. Spatially the plume extends sunward in the dusk sector of the dayside magnetosphere (Chen and Moore, 2006; Borovsky and Denton, 2008; Darrouzet et al., 2008), introducing a dawn–dusk asymmetry in the mass loading of the dayside outer magnetosphere. The effect of this asymmetry on solar-wind–magnetosphere coupling is discussed in Sect. 3.1.

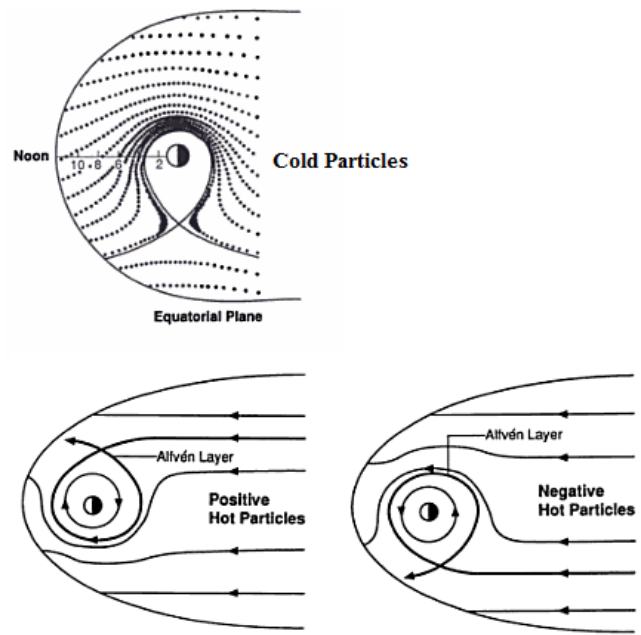

**Figure 12.** Drift paths of cold magnetospheric particles (top) and hot ions and electrons (bottom). The stable plasmasphere (closed drift paths inside of the separatrix) is shifted towards dusk.

### 2.5.4 Inner magnetosphere wave populations

Inner magnetospheric wave populations also exhibit dawn–dusk asymmetries. The spatial distribution of some inner magnetosphere wave populations is illustrated in Fig. 13, reproduced from Thorne (2010). Whistler mode chorus waves (Tsurutani and Smith, 1974) are typically found on the dawn side of the magnetosphere (Li et al., 2009) just outside the plasmapause and are linked to cyclotron resonant excitation of injected plasma sheet electrons (Li et al., 2008b). Thus the dawn–dusk asymmetry can be explained by considering the drift paths of the injected electrons (see Sects. 2.4.4 and 2.5.2). Electrostatic electron cyclotron harmonic waves are also linked to the injection of plasma sheet electrons into the inner magnetosphere (Horne and Thorne, 2000) and have a similar spatial distribution (Meredith et al., 2009). Plasmaspheric hiss is another whistler-mode emission that is mostly observed within the plasmasphere. Hiss also exhibits a dawn–dusk asymmetry: while average amplitudes of hiss are strongest on the dayside, emission extends into the pre-midnight sector at higher amplitudes than those observed in the post-midnight sector (Meredith et al., 2004). The generation of plasmaspheric hiss has recently been linked to the presence of chorus waves (Chum and Santolík, 2005; Bortnik et al., 2008; Bortnik et al., 2009), so one might expect them to have the same asymmetry. However, ray-tracing simulations have suggested that chorus-mode waves that are generated on the dayside can propagate eastwards and generate hiss in the dusk sector (Chen et al., 2009).

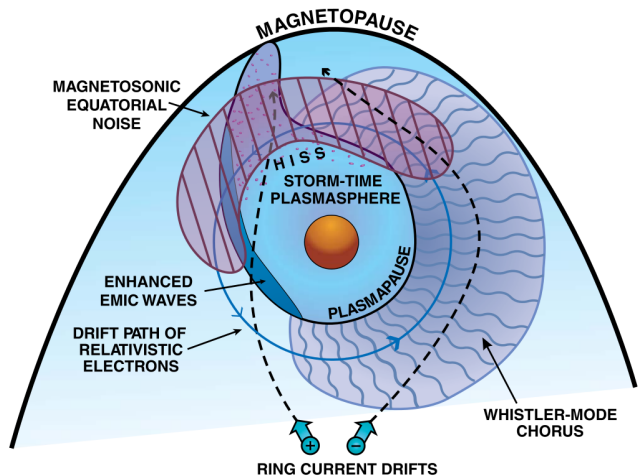

**Figure 13.** The spatial distribution of various inner magnetosphere wave populations (after Thorne, 2010).

Electromagnetic ion cyclotron (EMIC) waves are excited as a result of temperature anisotropy in ring current ions and also exhibit a dawn–dusk asymmetry. They typically occur in two frequency bands, just below the hydrogen and helium gyrofrequencies, respectively. The helium band waves dominate at dusk and are found between 8 and 12 $R_E$ whereas at dawn the hydrogen band waves dominate and are observed between 10 and 12 $R_E$ (Anderson et al., 1992; Min et al., 2012). EMIC wave power is typically larger at dusk than dawn (Min et al., 2012). EMIC waves have also been observed in the plasmaspheric plumes in the afternoon sector (Morley et al., 2009). Plumes can extend over a wide range of L-shells, and wave–particle interactions within them have been suggested as a source of asymmetric precipitation of ring current and radiation belt particles (Borovsky and Denton, 2009). While EMIC waves may scatter energetic particles during individual storms (e.g. Yuan et al., 2012), statistically EMIC waves are present only 10 % of the time in plasmaspheric plumes (Usanova et al., 2013).

Equatorial magnetosonic waves are another class of whistler-mode emission that are strongly confined to the equatorial plane. They have frequencies partway between the proton gyrofrequency and the lower hybrid frequency (e.g. Santolík et al., 2004). Equatorial magnetosonic waves have been observed both within and outside the plasmapause. Inside the plasmapause they are most intense at dusk (Meredith et al., 2008). Outside the plasmapause they are strongest in the dawn sector (Ma et al., 2013).

The spatial distribution of the whistler-mode chorus wave shown in Fig. 13 can be compared with the DMSP observations of diffuse aurora electron precipitation in Fig. 14 (top) (after Wing et al., 2013). The diffuse electron aurora has a strong dawn–dusk asymmetry and can be observed mainly between 22:00 and 10:00 MLT. As the plasma sheet electrons $E \times B$ convect earthward, they also curvature and gradient drift eastward toward dawn. The field-aligned component of these electrons is quickly lost through the loss cone, but they are replenished by pitch-angle scattering. A leading mechanism for pitch-angle scattering is very low frequency (VLF) whistler-mode chorus wave and electron interactions (e.g. Thorne, 2010; Reeves et al., 2009; Summers et al., 1998). Studies have shown that whistler-mode chorus waves are excited in the region spanning pre-midnight to noon. At around 10:00 MLT the diffuse electron flux decreases, which may suggest that the whistler-mode chorus waves start weakening. In the magnetosphere, the electrons continue to drift eastward, circling the Earth, but they are only observed in the ionosphere when and where there are whistler-mode chorus waves to pitch-angle scatter them. Contrast this with the asymmetry in monoenergetic auroral precipitation (Fig. 14, bottom) which peaks in the pre-midnight sector. This distribution will be discussed in more detail in Sect. 3.2.

## 2.6 Asymmetries in the thermosphere and ionosphere

The ionosphere has often been regarded as a projection of magnetospheric processes that are, in turn, driven by the solar wind, with the aurora as the most prominent manifestation. However, the ionosphere and its dawn–dusk asymmetries in particular can also have an impact on the magnetosphere. It is also important to bear in mind that in the thermosphere, up to approximately 1000 km altitude, the neutral density is still significantly higher than the ion density. Collisions between ions and neutrals cause exchange of momentum between the two species, so motion and dynamics of ions and neutrals influence each other.

Below, we show examples of dawn–dusk asymmetry in both neutrals and ions of the thermosphere and its embedded ionosphere.

### 2.6.1 The neutral atmosphere

In the thermosphere, i.e. the altitude range from approximately 85 up to 600 km, the dynamics are mainly dominated by dayside solar heating which drives a diurnal circulation of neutrals from the dayside to the nightside (e.g. Rees, 1979; Manson et al., 2002). Due to a combination of the Earth's rotation (which introduces an opposite effect of the Coriolis force at dawn and dusk) and the fairly slow transport, the induced noon–midnight asymmetry in neutral density and temperature becomes shifted towards a dawn–dusk asymmetry.

Figure 15 reproduced from Kervalishvili and Lühr (2013) shows maps of the relative thermospheric mass density enhancements ($\sigma_{rel} = \sigma / \sigma_{model}$) for three local seasons: winter, combined equinoxes and summer (measurements from Northern and Southern Hemisphere are combined). The dawn–dusk density asymmetry is most pronounced during local winter, when the solar illumination is minimum and the transport slower. Asymmetries in the neutral population also affect the ionosphere: due to collisions between neutrals and

**Table 5.** Overview of some pronounced dawn–dusk asymmetries in the inner magnetosphere.

| Process/property | Asymmetry preference | Source | References |
|---|---|---|---|
| Horizontal component of Earth's *B* field | stronger perturbation at dusk stronger perturbation at dusk | Ground magnetometers GOES, Polar, and Geotail | Chapman (1918); Newell and Gjerloev (2012) Tsyganenko et al. (2003) |
| Adiabatic heating of ring current ions | stronger at dusk | ENA simulations based on AMPTE/CCE/CHEM data | Milillo et al. (1996) |
| Ion distribution | peak at dawn | HENA | Fok et al. (2003) |

## Diffuse aurora electron precipitation

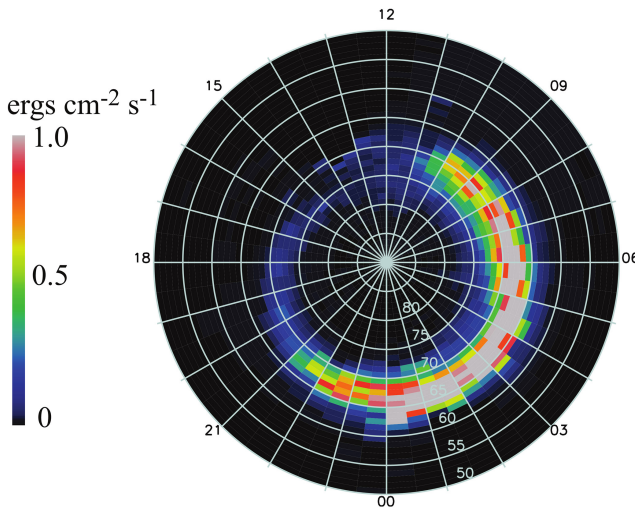

## Monoenergetic electron energy flux

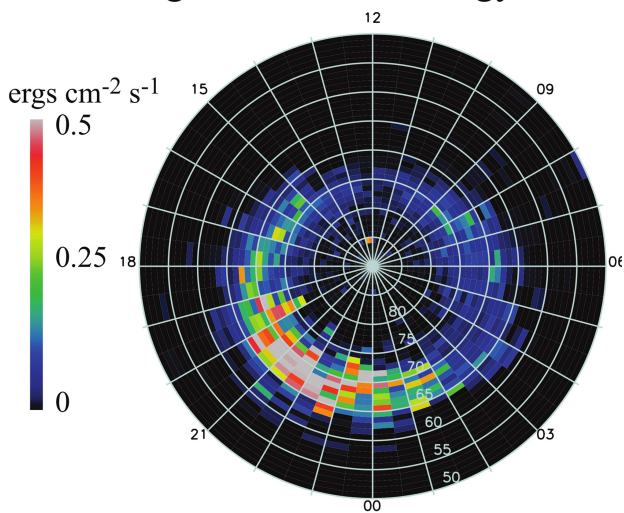

**Figure 14.** The spatial distribution of electron precipitation responsible for the diffuse aurora (top) and monoenergetic aurora (bottom). Note the different sense in the asymmetry of auroral emission (after Wing et al., 2013).

ions, a higher neutral density causes enhanced drag and thus reduced plasma convection (e.g. Förster et al., 2008). Also, higher neutral densities, as shown in Fig. 15, shift production levels of $O^+$ to higher altitudes, where reactions with other constituents such as $O_2$ and $NO_2$ are less frequent, thus increasing the escape probability. A comprehensive discussion about the interaction between the neutral atmosphere and the ionosphere is given in Bösinger et al. (2013)

### 2.6.2 Ionospheric convection

Embedded in the thermosphere is the ionosphere, with the highest ion concentrations around 200–400 km (the ionospheric F layer) where solar ultraviolet radiation (10–100 nm wavelength) induced ionisation of atomic and molecular oxygen is the dominant formation process.

The ionosphere is magnetically coupled to the magnetosphere, and the interaction between the solar wind with the dayside magnetopause will therefore also directly affect ionospheric convection. In particular, during a southward oriented IMF, a large-scale fast circulation of plasma in the magnetosphere is set up (Dungey, 1961). In the polar ionosphere, this circulation is manifested as two large-scale convection vortices. A cross-polar electric field is set up between the foci of the two vortices. Since this electric field is essentially the projection of the solar wind electric field across the reconnection line on the dayside, this cross-polar potential is often used as a proxy solar wind input energy to the magnetosphere.

Figure 16 shows maps of ionospheric convection in the Northern Hemisphere, in the form of potential plots. These synoptic maps were constructed from electric field measurements from the Cluster Electron Drift Instrument (EDI – see Paschmann et al., 2001) mapped down to 400 km altitude in the ionosphere, and converted to electric potentials by using the relation $E = -\nabla\Phi$. Ground-based studies based on the Super Dual Auroral Radar Network (SuperDARN – see Greenwald et al., 1995) give similar results. Southern Hemisphere patterns are similar, but are essentially mirrored with respect to dawn and dusk.

For purely southward IMF conditions (middle panel), the two large-scale convection cells are clearly apparent. The flow is mainly antisunward across the central polar cap, but skewed towards the pre-midnight sector behind the terminator. The dawn–dusk asymmetry is perhaps best seen in the left and right panels. A southward IMF with a positive IMF $B_Y$ component (right panel) rotates the convection cell patterns, and the main flow channel goes from around 10:00 MLT on the dayside to 21:00–22:00 MLT on the nightside. A negative IMF $B_Y$ of similar magnitude (left panel), however, leads to an almost straight noon–midnight plasma flow across the polar cap.

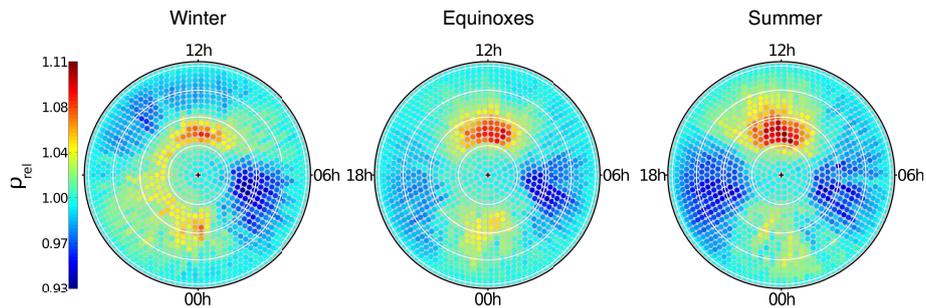

**Figure 15.** Colour-coded maps of neutral mass density anomalies in the thermosphere as measured by the Challenging Mini Payload (CHAMP) satellite at around 400 km altitude. Concentric circles indicate 50, 60, 70 and 80° magnetic latitudes. Left: local winter condition, i.e. minimum solar illumination. Middle panel, combined equinoxes measurements; right, summer conditions with maximum solar illumination. In particular during winter conditions a clear dawn–dusk asymmetry can be seen. After Kervalishvili and Lühr (2013). .

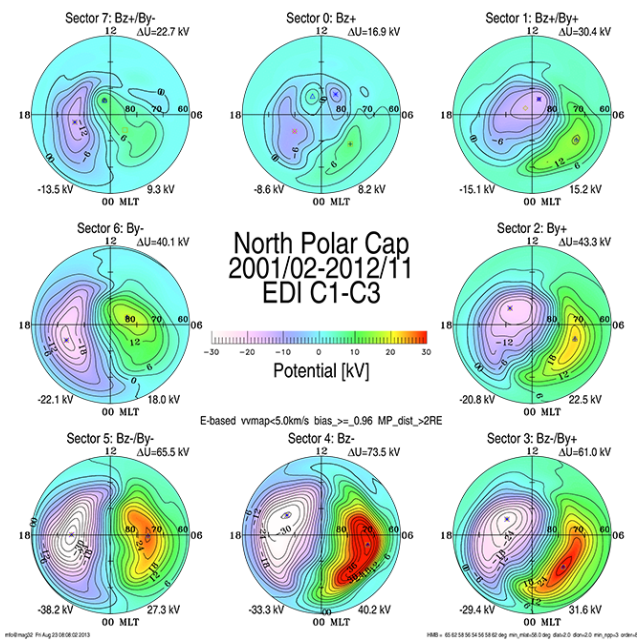

**Figure 16.** Maps of ionospheric convection in the Northern Hemisphere during southward IMF conditions. In the middle panel, the IMF is purely southward (i.e. clock angles around $180 \pm 22.5°$). The left and right panels show the influence of an additional IMF $B_Y$ component (left: clock angles $225 \pm 22.5°$, right: $135 \pm 22.5°$). The labels 12, 06, 00, 18 indicate magnetic local times; the concentric rings indicate 60, 70 and 80 degrees geomagnetic latitude. Plasma flows along equipotential lines; density of lines gives an indication of flow velocity. After Haaland et al. (2007). .

It is hard to envisage magnetospheric processes as the only source of these asymmetries. Atkinson and Hutchison (1978) attributed the lack of mirror symmetry to nonuniformities in ionospheric conductivity. They noted that a steep conductivity gradient across the day–night terminator tended to give a stronger squeezing of the plasma flow toward the dawnside of the polar cap. Tanaka (2001) used simulations with a realistic conductivity distribution to reproduce the observed asymmetries, and also noted that a uniform conductivity yielded symmetric convection cells.

The fact that the dawn–dusk mirror symmetry breaking can be explained by nonuniformities in ionospheric conductivity implies that magnetospheric convection is not simply the result of processes at the magnetospheric boundaries or in the magnetotail, but that it is modified by ionospheric effects.

### 2.6.3 Ionospheric outflow

Yau et al. (1984) found that upflow of both O$^+$ and H$^+$ with energies of 0.01 to 1 keV and pitch angles of 100–160° was larger at dusk. They also found a minimum in outflow in the post-midnight sector. They also noted that the asymmetry was altitude related, which they attributed to ion conic or beam acceleration. In a study by Pollock et al. (1990), however, the density of upwelling ions with low energies (0–50 eV/q) was found to have only a weak relation with magnetic local time, whereas the upwelling velocities differed for different ion species. Even with no asymmetry in the ionospheric source, transport of ionospheric plasma can cause asymmetric deposition in the magnetosphere. For example, Howarth and Yau (2008) used Akebono measurements to study trajectories of polar wind ions. They found a strong IMF $B_Y$ dependence, with deposition primarily in the dusk sector of the plasma sheet when IMF $B_Y$ was positive, and a more even distribution when IMF $B_Y$ was negative. Their study also suggested that ions emanating from the noon–dusk sector of the ionosphere could travel further in the tail, since the magnetic field lines are more curved. Likewise, Liao et al. (2010) examined the transport of O$^+$ (mainly from the cusp region) to the tail lobes. For IMF $B_Y$ positive, O$^+$ from the Northern Hemisphere cusp was found to be more likely to be transported to the dawn lobe, whereas O$^+$ from the Southern Hemisphere cusp/cleft region was transported to dusk.

The IMF $B_Y$-induced asymmetry and opposite effects for Northern Hemisphere and Southern Hemisphere can probably be explained by corresponding asymmetries in the dayside reconnection. This, again, leads to an asymmetric convection for the hemispheres (e.g. Haaland et al., 2007) and consequently in the transport of cold plasma from the ionosphere via the tail lobes to the plasma sheet.

In addition to the IMF $B_Y$-induced asymmetries, observations also indicate the presence of a persistent dawn–dusk asymmetry in plasma transport. Both Noda et al. (2003) and Haaland et al. (2008) noted a persistent duskward convection, unrelated to IMF direction. In Haaland et al. (2008) this asymmetry was related to the above-mentioned day–night conductivity gradient in the ionosphere (see Sect. 2.6.2). Furthermore, Yau et al. (2012) extended the single-particle simulation for the O$^+$ outflow in storm cases and found a clear dawn–dusk asymmetry. During five geomagnetic storms investigated, they found that the deposition of O$^+$ was on average $\sim$ 3 times higher in dusk than dawn plasma sheet.

A similar result, but using cold ion outflow (mainly protons with thermal and kinetic energy lower than 70 eV), was reported by Li et al. (2013). Figure 17, from this study, illustrates the persistent asymmetry. There is a larger deposition of cold ions of ionospheric origin in the dusk sector. In addition, there is also a strong IMF $B_Y$ modulation (not shown). Using the same data set, Li et al. (2012) also determined the source area for the cold ions, and found the polar cap regions to be the dominant contributors of cold plasma. Interestingly, no significant dawn–dusk asymmetry was found in the source.

## 3 Coupling between regimes

### 3.1 Solar wind – magnetosphere coupling

The impact of the solar wind on the Earth's magnetosphere drives activity in the magnetospheric system. The most significant coupling of the solar wind to the magnetosphere is via reconnection. While reconnection itself is most efficient under southward IMF $B_Z$, the orientation of the IMF $B_Y$ strongly influences asymmetries in the reconnection process. For a given event, a non-zero IMF $B_Y$ will result in many asymmetric signatures in the magnetosphere and ionosphere, by imposing a torque on the magnetic field flux tubes and their transport from dayside to nightside (Cowley, 1981). Such a torque leads to tail flux asymmetry and shifted nightside reconnection, and therefore asymmetries in particle populations and plasma convection in the plasma sheet. The lobes of the magnetosphere also experience density asymmetries under non-zero IMF $B_Y$, with the northern lobe having higher dawnside density under IMF + $B_Y$. The IMF $B_Y$ field penetrates to geosynchronous orbit, creating an asymmetry in geosynchronous $B_Y$ of 30 % (Cowley et al., 1983). The twisted open flux tubes also result in skewed ionospheric convection patterns (Ruohoniemi and Greenwald, 2005; Haaland et al., 2007, see also Fig. 16).

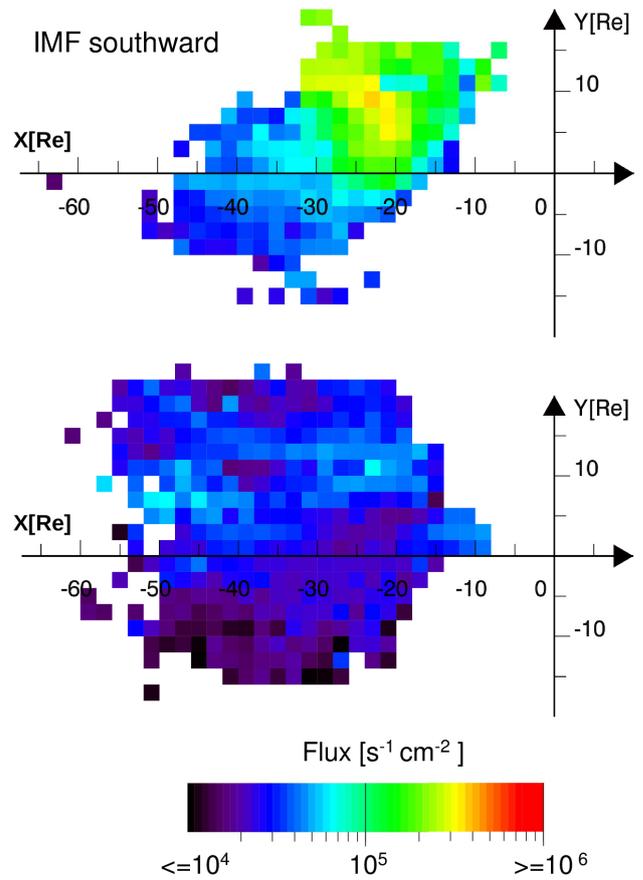

**Figure 17.** Maps of the deposition of cold ion flux from the ionosphere to the plasma sheet during periods with southward IMF conditions. The top panel shows the deposition of cold ions traced back to Cluster observations in the Northern Hemisphere polar cap and lobes, the lower panels shows the corresponding maps of ions traced back to the Southern Hemisphere. There is a clear dawn–dusk asymmetry with a higher fluxes, and thus larger deposition in the dusk sector. Adopted from Li et al. (2013).

Even when large statistical studies are used with average IMF $B_Y = 0$, many dawn–dusk asymmetries remain. IMF data are usually presented in the geocentric solar ecliptic (GSE) or the geocentric solar magnetospheric (GSM) systems, where the $x$ axis is defined as pointing from the Earth toward the Sun. The large majority of magnetospheric studies are presented in such coordinate systems. They are useful for displaying satellite trajectories, solar wind velocity and magnetic field measurements, magnetopause and bow shock positions, magnetosheath and magnetotail magnetic fields and plasma flows, etc. A solar wind velocity flowing straight from the Sun to the Earth would only have a $V_X$ component in such a system, with $V_Y = V_Z = 0$. However, this does not take into account the aberration, or rotation, of the solar wind due to the Earth's motion through space orbiting the Sun. Since the Earth is moving in the $-Y_{GSE}$ direction, a small rotation of the coordinate system is required to identify the true flow direction impacting on the Earth's magnetopause. The aberrated GSE coordinate system (AGSE) removes this small bias with the rotation angle $\theta_{aberr} = \tan^{-1}(V_E/V_{sw})$ where $V_E$ is the velocity of the Earth around the Sun (30 km s$^{-1}$). Many studies that present dawn–dusk asymmetries do not utilise the AGSE or AGSM coordinate systems.

**Table 6.** Ionospheric and thermospheric dawn–dusk asymmetries.

| Process/property | Asymmetry | Explanation | Reference |
|---|---|---|---|
| Large-scale convection | clockwise rotation of convection cells | ionospheric conductivity | Atkinson and Hutchison (1978); Tanaka (2001) Ridley et al. (2004) |
| Thermospheric density anomaly | higher densities on dusk | solar illumination, local heating, transport | Kervalishvili and Lühr (2013) |
| Coriolis force | opposing ion drift on dawn, enhancing on dusk | | Kervalishvili and Lühr (2013) |

Magnetosheath asymmetries are a direct result of solar wind driving. The motion of dayside reconnected flux tubes is asymmetric based on the IMF direction (Cooling et al., 2001) such that the IMF clock angle controls the location of flux transport event (FTE) signatures (Fear et al., 2012). In general, more FTEs are observed on the dusk sector of the magnetopause. Initially, this was attributed to stronger dusk-side magnetic field in the magnetosheath due to Parker spiral IMF draping (Kawano and Russell, 1996). However, recent results found that the differences in FTE occurrence by IMF spiral angle sector are not consistent with the Parker spiral IMF orientation (Y. L. Wang et al., 2006).

The magnetopause boundary becomes more asymmetric under strongly driven southward IMF $B_Z$, such that geosynchronous spacecraft are more likely to encounter the magnetopause on the dawn side rather than the duskside. Dmitriev et al. (2004) suggested that this could be due to either more intensive magnetopause erosion on the pre-noon/dawn sector, or the asymmetric ring current effect "pushing" the duskside magnetopause farther out. While the asymmetric ring current during storms is a result of ion drift toward dusk, solar wind pressure enhancements can increase the asymmetry of an already asymmetric ring current by inducing an azimuthal electric field that locally energises particles (Shi et al., 2005).

The coupling does not only operate in one direction; magnetospheric conditions can also change the solar-wind–magnetosphere coupling. Borovsky and Denton (2006) have proposed that the plasmaspheric plume will decrease solar-wind magnetospheric coupling or the geoeffectiveness of solar wind structures. When a plume extends to the magnetopause (Elphic et al., 1996; McFadden et al., 2008; B. M. Walsh et al., 2013) it will mass load a spatial region at the magnetopause, typically on the dayside. As the density increases, the localised reconnection rate will decrease causing a decrease in coupling (Borovsky et al., 2008). It is uncertain whether this localised decrease can be significant enough to impact the magnetospheric convection system.

### 3.2 Magnetosphere–ionosphere coupling

The ionosphere plays an active role in determining the state of magnetospheric convection, providing closure for the magnetospheric currents. The amount of current that can be carried through the ionosphere is determined by ionospheric conductivity. It has been noticed that the day–night gradient of the ionospheric conductivity produces the dawn–dusk asymmetry in the polar cap convection (Atkinson and Hutchison, 1978). Observations and modelling suggest that the two-cell ionospheric convection pattern is rotated clockwise with respect to the noon–midnight meridian even for IMF $B_Y \simeq 0$ conditions (e.g. Ridley et al., 2004; Ruohoniemi and Greenwald, 2005; Haaland et al., 2007; Cousins and Shepherd, 2010, see also Sect. 2.6.2 and Fig. 16).

The dawn–dusk asymmetry in ionospheric convection resulting from the conductance gradient (e.g. Atkinson and Hutchison, 1978; Tanaka, 2001; Ridley et al., 2004) may affect the geometry of magnetotail lobes and, therefore, the geometry of plasma and current sheet. Zhang et al. (2012) use three-dimensional global MHD Lyon–Fedder–Mobarry (LFM) model to simulate a magnetosphere response on solar wind/IMF driving. The realistic model of the ionospheric conductance included effects of electron precipitation and solar UV ionisation. The numerical experiment was controlled to eliminate all asymmetries and variability in the solar wind to isolate an effect of the ionospheric state on magnetotail activity. These controlled simulations by Zhang et al. (2012) suggest that the ionospheric conductance can regulate the distribution of fast flows in the magnetotail so that the flows are more intense in the pre-midnight plasma sheet.

The simulations by Zhang et al. (2012) have revealed that gradients in Hall ionospheric conductance are necessary to create the dawn–dusk asymmetry (note that neither IMF $B_Y$ nor solar wind $V_Y$ were included). These simulations are confirmed by observations; the observed distributions of Hall conductance lead to a rotation in the polar cap convection in order to preserve current continuity. The rotation results in the displacement of the symmetry axis of the two-cell convection from the noon–midnight meridian to the 11:00–23:00 LT as shown in Fig. 16. The clockwise rotation of the convection pattern causes more open flux to be diverted towards the duskside of the magnetotail. This results in dawn–dusk asymmetry of loading and, consequently, reconnection of magnetic flux in the plasma sheet (Smith, 2012). Numerical tests including clockwise as well as (unrealistic) anticlockwise rotation of the polar cap convection pattern have shown a linear correlation between a degree of convection pattern rotation and a degree of reconnection asymmetry.

The ionospheric outflow may also influence the processes in the magnetotail plasma sheet. It has been argued by Baker et al. (1982) that asymmetries in the distribution of enhanced density of $O^+$ may define regions in the plasma sheet where tearing mode growth rate are increased and the instability threshold is lowered. They pointed out that statistical studies of $O^+$ concentration in the plasma sheet revealed significant dawn–dusk asymmetry with larger occurrence rate in the pre-midnight sector. Adopting the criterion for onset of the linear ion tearing instability (Schindler, 1974), Baker et al. (1982) studied the possible role of the ionospheric $O^+$ ions in the development of plasma sheet tearing. Their analysis resulted in maximum tearing growth rate in the range of $-15 < X_{GSM} < -10\,R_E$ and $Y_{GSM} \sim 5\,R_E$. Recent statistical studies of Geotail/EPIC data have confirmed that average energy of the $O^+$ ions increases toward dusk (Ohtani et al., 2011).

The observed asymmetry in monoenergetic auroral electron precipitation (Fig. 14, bottom) is also thought, in part, to be a result of magnetosphere–ionosphere coupling. The precipitating energy flux can be associated with the upward region 1 field-aligned currents, which are mostly located in the pre-midnight sector (e.g. Wing et al., 2013, and references therein).

### 3.2.1 Plasma sheet and the inner magnetosphere

As geomagnetic activity increases, the boundary between open and closed drift paths moves closer to Earth. Thus, protons and electrons from the plasma sheet are able to access geosynchronous orbit during storms. Using LANL-MPA (Los Alamos National Laboratory Magnetospheric Plasma Analyzer) measurements, Korth et al. (1999) found higher densities toward dawn for both electrons and ions (with energies 1 eV–40 keV) at geosynchronous orbit during periods of higher geomagnetic activity. For low geomagnetic activity, the electron and ion densities peak at midnight, but the reasons for lower densities at dawn and dusk differ. For electrons, the duskside region is dominated by closed drift paths for electron plasma sheet energies while plasma sheet electrons are lost to precipitation on the dawn side. For protons, the ions take longer to drift toward the duskside, allowing more losses to precipitation. Temperatures also exhibit an asymmetry – with hotter ion temperatures toward dusk. In addition to the gradient–curvature drift yielding higher ion temperatures toward dusk in the magnetotail, higher energy ions that drift toward dawn are preferentially lost to particle precipitation (Denton et al., 2006). During a geomagnetic storm, ion temperatures toward dusk increase while those toward dawn decrease, yielding a more pronounced asymmetry around minimum Dst. Such cold temperatures in the dawn–noon sector have been observed during geomagnetic storms with in situ measurements at geosynchronous orbit (Denton et al., 2006) and with remote TWINS ENA measurements (Keesee et al., 2012).

During enhanced geomagnetic activity, plasma sheet ions penetrate deep into the inner magnetosphere (e.g. Ganushkina et al., 2000; Runov et al., 2008). The low-energy (< 10 keV) part of this population is subject to the co-rotation drift and drifts dawnward, whereas the high-energy (> 10 keV) part drifts duskward following gradient- and curvature-drift paths (see Fig. 12). A population with energy ∼ 10 keV often becomes "stagnant", forming the so-called "ion nose structures" because of a characteristic shape of the energy spectrogram (e.g. Ganushkina et al., 2000). Statistical studies of ion nose structures observed by Polar/CAMMICE revealed dawn–dusk asymmetry in the event distribution with larger occurrence rate in the dusk sector.

In general, enhanced plasma sheet convection and energetic plasma sheet particle injections build up an asymmetric pressure in the inner magnetosphere with stronger enhancement on the duskside that results from asymmetric drifts of energetic ions and electrons. Duskward gradient and curvature drifts of energetic ions lead to localised pressure increases.

## 4 Open Issues and inconsistencies

Many of the dawn–dusk asymmetries discussed in the previous sections can be explained by asymmetries in the input. In particular, the IMF interaction with the magnetosphere is known to impose significant asymmetries in the plasma entry and flux transport. On the other hand, the difference in behaviour/motion of ions and electrons in nonuniform fields is another source of asymmetries. However the relative importance of these two mechanisms is largely unknown.

Below, we try to identify some still-open issues in our understanding of dawn–dusk asymmetries observed in the Earth's magnetosphere and ionosphere.

### 4.1 External versus internal influence

As seen in Sects. 2.1 and 2.1.2, pronounced dawn–dusk asymmetries exist in the magnetosheath. A still open question is the degree to which this asymmetry translates into a corresponding asymmetry inside the magnetopause, and whether this can explain e.g. the observed asymmetries in observed properties and processes in the nightside plasma sheet.

The relative importance of the ionosphere for magnetospheric dawn–dusk asymmetries is also largely unknown. Conductivity effects as discussed in Sects. 2.6.2 and 3.1 are believed to cause a local ionospheric asymmetry in the ionospheric plasma transport, but their effect on magnetotail flows is still disputed. Likewise, neutral density and wind can influence both ion outflow and ionospheric drag, but the role of the thermosphere for large-scale magnetospheric dawn–dusk asymmetries is still largely unknown.

## 4.2 Ring current closure

One of the first scientific observations of a dawn–dusk asymmetry in geospace was reported by Chapman (1918). He noted that ground magnetic perturbations associated with geomagnetic storms were larger at dusk. The first direct observations of an asymmetric ring current were made in the early 1970s (e.g. Frank, 1970) as spacecraft observations became available. An asymmetry in the ring current naturally raises the question of current closure. Initially, the observed dawn–dusk asymmetry, or partial ring current, was mainly attributed to divergence either through field-aligned currents into the ionosphere, through the cross-tail current or as local current loops within the magnetosphere (e.g. Liemohn et al., 2013). The recent results from Haaland and Gjerloev (2013) indicate a mutual influence between the ring current and magnetopause current, although a clear current loop connecting the ring current with the magnetopause current has not been firmly established.

## 4.3 The impact of the plume on magnetospheric driving

As discussed in Sect. 2.5 the plasmaspheric plume is capable of transporting large amounts of plasma from the dense plasmasphere to the outer magnetosphere, primarily in the dusk sector. Mass loading of the dayside magnetopause in this region has been shown to impact reconnection (B. M. Walsh et al., 2013) and could impact the efficiency of solar-wind–magnetosphere coupling. Borovsky et al. (2013) predict that the plume can reduce reconnection by up to 55 % during coronal mass ejections (CMEs) or high-speed streams. On a larger scale, Borovsky and Denton (2006) looked at geomagnetic activity with and without a plume present at geosynchronous orbit and concluded that the impact of the plume is significant enough to reduce geomagnetic activity.

By contrast, Lopez et al. (2010) argue that although the plume may reduce the reconnection rate locally where high-density material contacts the magnetopause, the total reconnection rate integrated across the full X-line should not change significantly. In the Lopez et al. (2010) model, the density asymmetry would not significantly impact magnetospheric convection or the development of storms. Understanding the full impact of the plume on reconnection and storm dynamics remains an open issue.

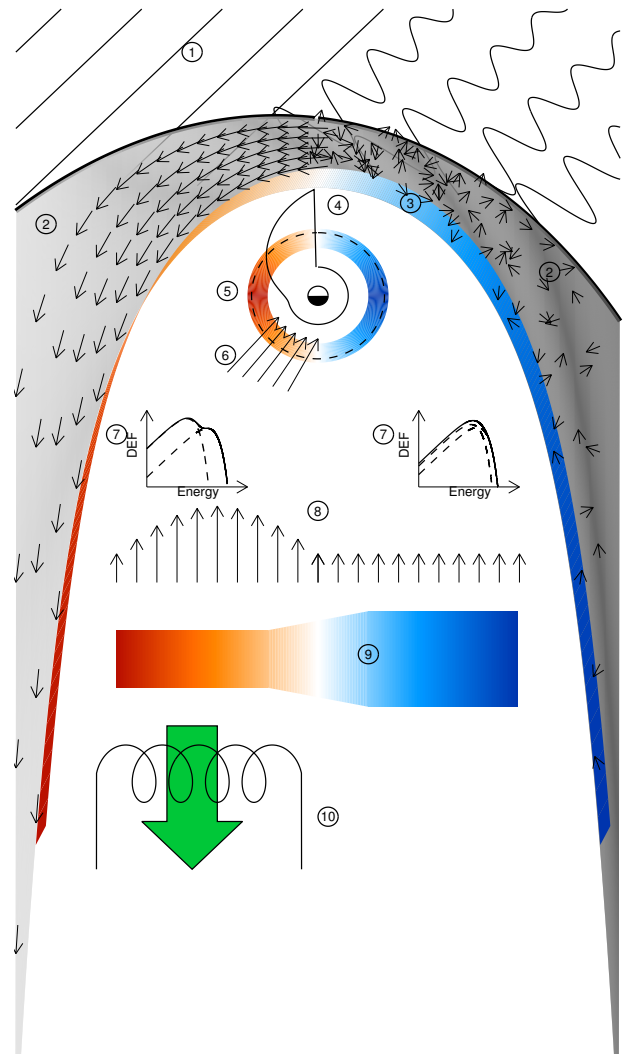

**Figure 18.** Schematic diagram showing some of the identified dawn–dusk asymmetries in the coupled solar wind–magnetosphere–ionosphere system. The configuration for a Parker spiral orientation of the IMF is shown. (1) The foreshock shows a greater occurrence of ULF waves in the quasi-perpendicular shock region towards dawn; (2) the magnetosheath is thinner, more turbulent and denser at dawn, but magnetic field strength is greater at dusk; (3) the magnetopause is thicker at dawn, but the magnetopause current density is greater at dusk; (4) the plasmasphere extends out to the magnetopause in plumes, typically seen on the duskside; (5) the ring current is asymmetric and stronger on the duskside; (6) high energy particle injections at geosynchronous orbit are more common on the duskside; (7) magnetotail ions are made up of hot and cold populations – the hot population is colder and the cold population is hotter towards dawn (distributions shown in differential energy flux); (8) the occurrence of convective fast flows in the tail shows no dawn–dusk asymmetry, but flows towards dusk are faster; (9) the magnetotail current sheet is thicker towards dawn and the current density is greater towards dusk; (10) signatures of reconnection are more commonly seen towards dusk.

## 5 Summary and conclusions

Asymmetries are ubiquitous features of the Earth's magnetosphere and plasma environment. Noon–midnight asymmetries are mainly governed by solar illumination resulting in strongly asymmetric ionisation in the nightside and dayside. Magnetic gradients due to the compressed sunward-facing magnetosphere on noon and the corresponding stretched magnetotail tail in the nightside also introduces a significant noon–midnight asymmetry. Similarly, north–south asymmetries can often be explained by seasonal differences in illumination of the two hemispheres, and consequently differences in ionospheric conductivity. Differences in the geomagnetic field between the two hemispheres will also create north–south asymmetries in ionospheric plasma motion.

Persistent dawn–dusk asymmetries, on the other hand, have received less attention and are not always easy to explain. In this paper, we have tried to give an overview of prominent dawn–dusk observational features and their possible explanations. Figure 18 gives a schematic overview of some of the dawn–dusk asymmetries discussed in this paper. We have focused on four key aspects: (1) the role of external influences such as the solar wind and its interaction with the Earth's magnetosphere; (2) properties of the magnetosphere itself; (3) the role of the ionosphere for magnetospheric dynamics, and (4) the coupling between the solar wind, magnetosphere and ionosphere.

As reviewed in Sect. 2.1, external factors such as bow shock geometry and direction of the interplanetary magnetic field, labelled (1) and (2) in Fig. 18, are important for dawn–dusk asymmetries. The shock geometry creates an asymmetry in plasma properties at dawn and dusk of the magnetosheath. In addition, the IMF orientation exerts significant control of both magnetospheric and ionospheric processes. A key element here is the dayside interaction between the IMF and the geomagnetic field, and IMF $B_Y$ is perhaps the strongest driver of dawn–dusk asymmetry in the magnetosphere. This interaction is also manifested in the ionosphere where the large-scale plasma convection pattern shows a systematic response to IMF orientation.

Asymmetries in the magnetosheath are also reflected inside the magnetosphere. In Sect. 2.3 we pointed out the role of plasma entry from the magnetosheath along the magnetopause flanks. Differences in dawn and dusk magnetosheath plasma properties will consequently influence geometry (9), plasma properties (7) and processes in the magnetotail (8), (10).

External drivers are not fully able to explain all dawn–dusk asymmetry, though. As discussed in Sect. 2.5, a noticeable dawn–dusk asymmetry arises as a consequence of gradient and curvature drift of particles; electrons and ions are deflected in opposite directions. This is most pronounced for the inner magnetosphere, where the magnetic gradients are stronger. A prominent example is the asymmetric ring current (5), with a stronger net current on the duskside.

In Sect. 2.6 we discussed dawn–dusk asymmetries in the thermosphere and its embedded ionosphere. In addition to asymmetries imposed by the magnetosphere, these regions also possess locally induced dawn–dusk asymmetries. Differences in thermospheric heating and conductivity gradients in the ionosphere are two prominent examples.

In order to fully understand the dynamic behaviour of geospace, including mechanisms responsible for dawn–dusk asymmetry, we must treat the solar wind, magnetosphere and ionosphere as a fully coupled system. As seen in Sect. 3, key aspects in regulating the response of this coupled system are the degree of feedback provided by the magnetosphere to the solar wind input, and the feedback from the ionosphere to the magnetosphere. The feedback from the ionosphere, both in the form of ion outflow (discussed in Sect. 2.6.3) and the role of ionospheric conductivity (discussed in Sect. 3.2) have been studied extensively, and are believed to influence the magnetosphere. Magnetospheric feedback to the magnetopause and bow shock regions, for example the effect of the plume (labelled (4) in Fig. 18) on dayside reconnection (discussed in Sect. 4.3) is still largely unexplored, however. It is therefore fair to say that there are still major gaps in our understanding of phenomena that introduce asymmetries in geospace.

*Acknowledgements.* This paper is the output of the International Space Science Institute international team "dawn–dusk asymmetry in the Coupled Solar Wind Magnetosphere Ionosphere system". The authors wish to acknowledge ISSI for funding our meetings in Berne. S. Haaland acknowledges support by the Norwegian Research Council under contract 223252/F50. A. M. Keesee acknowledges support from NASA EPSCoR Award NNX10AN08A and WVU Research and Scholarship Committee. J. Soucek acknowledges the support of grant P209/12/2394 of Czech Science Foundation GACR. S. Wing acknowledges the support of NSF grant AGS-1058456. C. Forsyth acknowledges support of STFC Consolidated Grant ST/K000977/1. J. Kissinger was supported by an appointment to the NASA Postdoctoral Program at the Goddard Space Flight Center, administered by Oak Ridge Associated Universities through a contract with NASA.

Topical Editor E. Roussos thanks R. A. Treumann and R. Fear for their help in evaluating this paper.